\documentclass[journal=jacsat,manuscript=article]{achemso}

\usepackage[version=3]{mhchem} % Formula subscripts using \ce{}
\usepackage{fancyvrb}
\usepackage{xcolor}
\usepackage[most]{tcolorbox}
\usepackage{setspace}
\usepackage{comment}

\author{Daniele Rapetti}
\affiliation[SISSA]{Scuola Internazionale Superiore di Studi Avanzati (SISSA), Via Bonomea 265, 34136 Trieste, Italy}
\author{Massimiliano Bonomi}
\affiliation{Institut Pasteur, Université Paris Cité, CNRS UMR 3528, Computational Structural Biology Unit, Paris, France}
\email{mbonomi@pasteur.fr}
\author{Carlo Camilloni}
\affiliation{Department of Biosciences, University of Milano, Milano, Italy}
\email{carlo.camilloni@unimi.it}
\author{Giovanni Bussi}
\email{bussi@sissa.it}
\affiliation[SISSA]{Scuola Internazionale Superiore di Studi Avanzati (SISSA), Via Bonomea 265, 34136 Trieste, Italy}
\author{Gareth A. Tribello}
\affiliation
{Centre for Quantum Materials and Technologies, School of Mathematics and Physics, Queen's University Belfast, Belfast, United Kingdom}
\email{g.tribello@qub.ac.uk}

\title[Making PLUMED fly]
  {Making PLUMED fly: a tutorial on optimizing performance}

\abbreviations{IR,NMR,UV}

\begin{document}

%%%%%%%%%%%%%%%%%%%%%%%%%%%%%%%%%%%%%%%%%%%%%%%%%%%%%%%%%%%%%%%%%%%%%
%% The "tocentry" environment can be used to create an entry for the
%% graphical table of contents. It is given here as some journals
%% require that it is printed as part of the abstract page. It will
%% be automatically moved as appropriate.
%%%%%%%%%%%%%%%%%%%%%%%%%%%%%%%%%%%%%%%%%%%%%%%%%%%%%%%%%%%%%%%%%%%%%
\begin{tocentry}

\includegraphics[width=7cm]{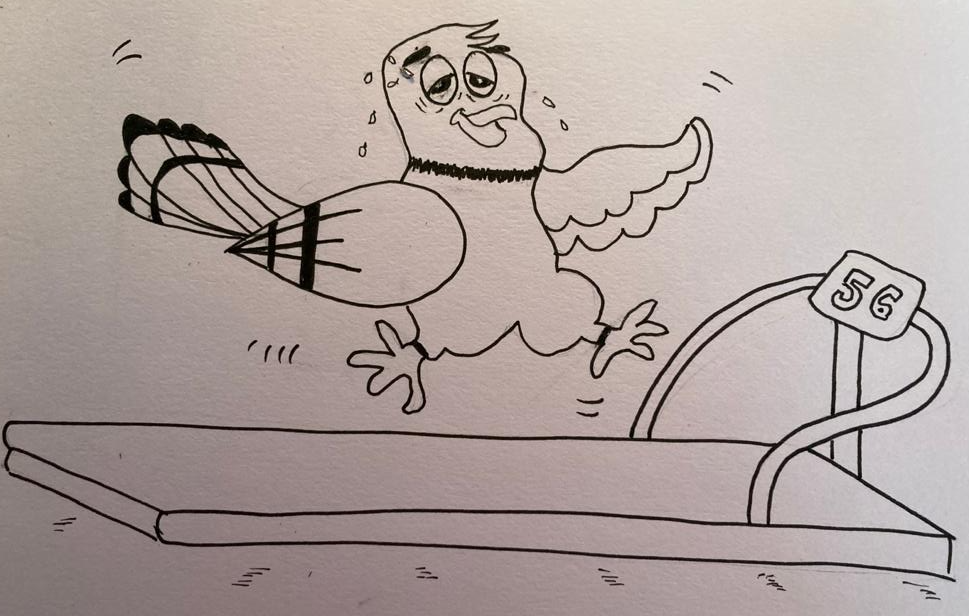}

\end{tocentry}

%%%%%%%%%%%%%%%%%%%%%%%%%%%%%%%%%%%%%%%%%%%%%%%%%%%%%%%%%%%%%%%%%%%%%
%% The abstract environment will automatically gobble the contents
%% if an abstract is not used by the target journal.
%%%%%%%%%%%%%%%%%%%%%%%%%%%%%%%%%%%%%%%%%%%%%%%%%%%%%%%%%%%%%%%%%%%%%
\begin{abstract}
PLUMED is an open-source software package that is widely used for analyzing and enhancing molecular dynamics simulations that works in conjunction with most available molecular dynamics softwares. While the computational cost of PLUMED calculations is typically negligible compared to the molecular dynamics code's force evaluation, the software is increasingly being employed for more computationally demanding tasks where performance optimization becomes critical.  In this tutorial, we describe a recently implemented tool that can be used to reliably measure code performance. We then use this tool to generate detailed performance benchmarks that show how calculations of large-numbers of distances, angles or torsions can be optimized by using vector-based commands rather than individual scalar operations.  We then present benchmarks that illustrate how to optimize calculations of atomic order parameters and secondary structure variables. Throughout the tutorial and in our implementations we endeavor to explain the algorithmic tricks that are being used to optimize the calculations so others can make use of these prescriptions both when they are using PLUMED and when they are writing their own codes.
\end{abstract}

%%%%%%%%%%%%%%%%%%%%%%%%%%%%%%%%%%%%%%%%%%%%%%%%%%%%%%%%%%%%%%%%%%%%%
%% Start the main part of the manuscript here.
%%%%%%%%%%%%%%%%%%%%%%%%%%%%%%%%%%%%%%%%%%%%%%%%%%%%%%%%%%%%%%%%%%%%%
\section{Introduction}

PLUMED (\url{https://www.plumed.org}) \cite{plumed} is an open-source software package that can be used to analyze and enhance molecular dynamics (MD) trajectories. Rather than operating as a monolithic software package, PLUMED serves as a framework 
%that provides voluntary standards and conventions 
%for researchers developing and sharing advanced sampling techniques.  
for researchers who are using and developing advanced sampling techniques to share ideas.
Consequently, in addition to the code, we also maintain a repository for sharing the inputs that have been used in publications that employ PLUMED (\url{https://www.plumed-nest.org}) \cite{plumed-nest} and a site for sharing tutorials that explain how to use its features (\url{https://www.plumed-tutorials.org}) \cite{plumed-tutorials}. 

\begin{figure}
    \centering
    \includegraphics[width=0.5\textwidth]{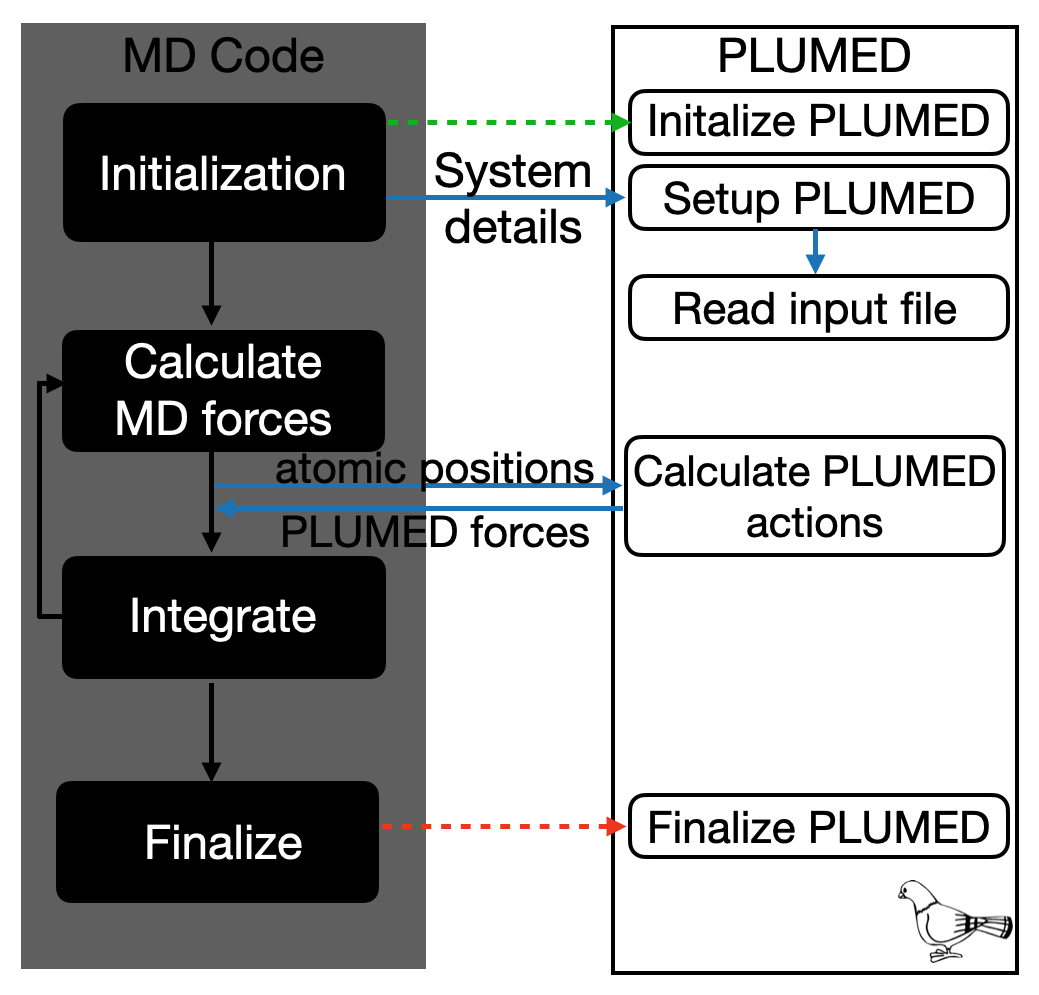}
    \caption{Interaction of PLUMED with molecular dynamics codes. PLUMED is able to calculate functions of the atomic positions and apply forces to atoms by passing data to and from the underlying MD code.}
    \label{fig:MD-interface}
\end{figure}

As shown in Fig. \ref{fig:MD-interface}, PLUMED is designed to be inserted into MD codes and to work alongside them. It works by receiving the atomic positions from the underlying MD code, calculating various quantities from these positions and, if necessary, adjusting the forces that are acting upon the atoms.  As we will discuss in the background section of this paper, PLUMED is written in a way that makes adding functionalities to calculate new quantities from these positions straightforward. This, combined with the PLUMED’s interoperability, has enabled developers to contribute their methodologies \cite{clusterpaper,mlcolvar,joaquin,baldi,bingqing3,trout,Tribello_dimred1,tribello-dimred2,drr_module,eds_module,envsim_module,fiist_module,funnel_module,isdb_module,logmfd_module,membrane_module,opes_module,pamm_module,piv_module,pytorch_module,pytorch_module2,s2cm_module,perovskite,sprint_module,ves_module,itre,atlas,10.1371/journal.pcbi.1012180,Schnapka2025.06.18.660298,shaman,hsu2023alchemical,frohlking2022automatic,cesari2016combining,bottaro2016free} within a unified ecosystem while maintaining compatibility across different molecular dynamics engines \cite{gromacs,lammps,openmm,metalwalls,plumed_gui,amber1,amber2,cp2k,qe1,qe2,qe3,dlpoly5,ipi,dftb,ase,gpumd,namd,ops1,ops2,htmd}. By establishing common interfaces and documentation standards, PLUMED has transformed how enhanced sampling methods are disseminated, validated, and adopted and effectively democratized access to cutting-edge simulation techniques and accelerated methodological innovation in the field.

Typically the computational cost of running PLUMED alongside an MD code is small. If PLUMED is being used to calculate a simple function of a small number of atomic positions, the cost of this calculation is negligible when compared to the cost associated with calculating the atomic forces.  However, PLUMED is increasingly being used to perform more computationally expensive calculations.  Given that in such calculations PLUMED can have a significant effect on performance, a tutorial paper that explains how to get the best performance from PLUMED feels timely.  In the following sections we will thus explain some of the more complicated calculations that PLUMED can be used to perform and will then discuss various approaches that can be used to make the parts of these calculations that are performed in PLUMED run more quickly.

\section{Background}

Input to PLUMED typically consists of a single file that provides instructions on what PLUMED should calculate.  An example input that illustrates how we can use PLUMED to add a restraint that forces the vector connecting atoms 1 and 2 to point in a direction that is perpendicular to the 111 direction of the lab frame is shown below. 
\begin{tcolorbox}[top=-10pt] 
\singlespacing \footnotesize 
\begin{Verbatim}[commandchars=\\\{\}] 
\noindent \textcolor{blue}{\# Calculate the vector connecting atoms 1 and 2} \\
{\bf d}: \textcolor{green}{DISTANCE} ATOMS=1,2 COMPONENTS \\
\textcolor{blue}{ \# Calculate the modulus of the vector connecting atom 1 and 2} \\ 
{\bf dm}: \textcolor{green}{CUSTOM} ARG=d.x,d.y,d.z FUNC=sqrt(x*x+y*y+z*z) PERIODIC=NO \\
\textcolor{blue}{ \# Calculate the director of the vector connecting atoms 1 and 2} \\
{\bf ux}: \textcolor{green}{CUSTOM} ARG=d.x,dm FUNC=x/y PERIODIC=NO  \\
{\bf uy}: \textcolor{green}{CUSTOM} ARG=d.y,dm FUNC=x/y PERIODIC=NO \\
{\bf uz}: \textcolor{green}{CUSTOM} ARG=d.z,dm FUNC=x/y PERIODIC=NO \\
\textcolor{blue}{ \# Find the dot product between the director of the vector connecting} \\
\textcolor{blue}{ \# atoms 1 and 2 and the 111 direction.} \\ 
{\bf cc}: \textcolor{green}{COMBINE} ARG=ux,uy,uz COEFFICIENTS=1,1,1 NORMALIZE PERIODIC=NO\\  
\textcolor{blue}{ \# Calculate the angle between the vector connecting atoms 1 and 2} \\ 
\textcolor{blue}{ \# and the 111 direction} \\
{\bf ang}: \textcolor{green}{CUSTOM} ARG=cc FUNC=acos(x) PERIODIC=NO \\
\textcolor{blue}{ \# Add an harmonic restraint to force this angle to remain close to 90 degrees} \\ 
{\bf res}: \textcolor{green}{RESTRAINT} ARG=ang AT=pi/2 KAPPA=100 \\
\end{Verbatim}
\end{tcolorbox}

\begin{figure}
\includegraphics[width=0.5\textwidth]{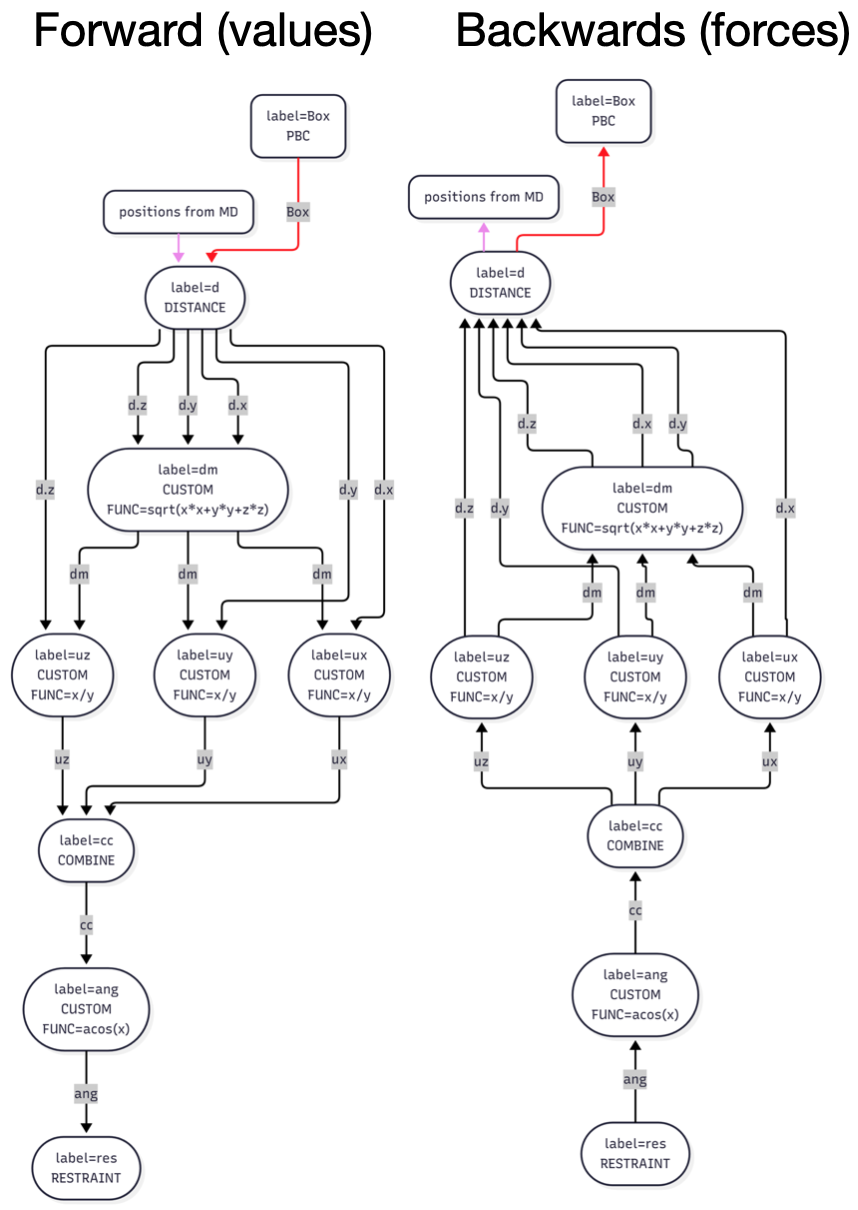}
\caption{Data communications between PLUMED actions. The left panel illustrates how the constituent actions in the first example input in this paper evaluate the bias function.  The right panel shows how data is passed between actions when forces are evaluated using the chain rule.  These figures were made by using the PLUMED command {\tt plumed~show\_graph}, which outputs a Mermaid diagram.}
\label{fig:first-graph}
\end{figure}

Although this input performs a calculation that users would likely not perform, it does nicely illustrate PLUMED’s philosophy. 
%This plugin code is written primarily by and for people working on research in academia. It is not designed to be a black box - we want users who fully comprehend the calculations they are performing.  We will thus often try expose the mathematical procedures that are used in calculations in the input syntax.
%As this is a code written by (and often for) people who develop new methodologies, we have an input that seeks to illustrate how calculations are performed.  
Each line in the input above defines an action and, by passing values between these actions, as illustrated in Fig. \ref{fig:first-graph} for the above input, the input defines a chain of operations that should be performed on the input positions. Notice that we also run through the action list backwards as illustrated in the right panel of Fig. \ref{fig:first-graph} in order to evaluate the forces that our actions apply on the atomic positions.
Also notice that, because the vector connecting the two atoms here is evaluated using the minimal image convention, it depends on the simulation box size. As such, when we do the backward propagation of the forces additional forces are added on the positions of the 
two atoms and on the cell vectors.  These forces on the cell vectors contribute to the internal pressure of the system.

The input above also illustrates how we routinely use PLUMED’s {\tt CUSTOM} action, which relies on the Lepton library \cite{lepton} that was developed by Peter Eastman and
that we extracted from OpenMM \cite{openmm}.  This action provides users with a way of specifying arbitrary functions to be applied on the input values. In the above input one can thus see how the director of the vector connecting atoms 1 and 2 is computed from the components of the unnormalized vector and how the angle between this vector and the $(111)$ direction is computed by calculating the arccosine of a dot product.

As indicated in the left panel of Fig. \ref{fig:first-graph}, to pass the quantities calculated by one action to another action that is defined later you use the labels that appear before the colon in the line of the first action in the input to the {\tt ARG} keyword for the second action. In the input above, the quantities that are passed between the actions are all scalars.  However, from PLUMED 2.10 onwards you can also pass vectors in the same way.  The following example  illustrates how this passing of vectors works in practice. To make clear the types of value begin passed we write the labels for quantities that are vectors in blue and labels for quantities that are scalars in black in this as well as all the inputs that follow.

\begin{tcolorbox}[top=-10pt]
\singlespacing \footnotesize
\begin{Verbatim}[commandchars=\\\{\}]
\noindent \textcolor{blue}{\# Calculate vectors that define the orientation of a collection of molecules}
\textcolor{blue}{\bf d}: \textcolor{green}{DISTANCE} ATOMS1=1,2 ATOMS2=3,4 ATOMS3=5,6 ATOMS4=7,8 \textcolor{blue}{# you can add more pairs here}
\textcolor{blue}{ # Now calculate the modulus of all the vectors above}
\textcolor{blue}{\bf dm}: \textcolor{green}{CUSTOM} ARG=\textcolor{blue}{d.x},\textcolor{blue}{d.y},\textcolor{blue}{d.z} FUNC=sqrt(x*x+y*y+z*z) PERIODIC=NO
\textcolor{blue}{ # Calculate the directors of all the vectors}
\textcolor{blue}{\bf ux}: \textcolor{green}{CUSTOM} ARG=\textcolor{blue}{d.x},\textcolor{blue}{dm} FUNC=x/y PERIODIC=NO 
\textcolor{blue}{\bf uy}: \textcolor{green}{CUSTOM} ARG=\textcolor{blue}{d.y},\textcolor{blue}{dm} FUNC=x/y PERIODIC=NO
\textcolor{blue}{\bf uz}: \textcolor{green}{CUSTOM} ARG=\textcolor{blue}{d.z},\textcolor{blue}{dm} FUNC=x/y PERIODIC=NO
\textcolor{blue}{ # Calculate the mean from the three vectors above}
{\bf mx}: \textcolor{green}{MEAN} ARG=\textcolor{blue}{ux} PERIODIC=NO
{\bf my}: \textcolor{green}{MEAN} ARG=\textcolor{blue}{uy} PERIODIC=NO
{\bf mz}: \textcolor{green}{MEAN} ARG=\textcolor{blue}{uz} PERIODIC=NO
\textcolor{blue}{ # And calculate the modulus of average}
{\bf cv}: \textcolor{green}{CUSTOM} ARG=ux,uy,uz FUNC=sqrt(x*x+y*y+z*z) PERIODIC=NO
\end{Verbatim}
\end{tcolorbox}

The final quantity calculated in this input file is an order parameter that has been used to study liquid crystals \cite{liquid_crystal}.  Each pair of atoms specified in the distance command of the input gives an orientation for one of the molecules in the liquid crystal.  The scalar quantity {\bf cv} is thus equal to one if all the molecules in the liquid crystal have the same orientation and zero if all these molecules all have wildly different orientations. 

To keep input files short, we provide shortcut commands.
%that PLUMED expands to longer inputs for many variables.  
So, for example, you can calculate {\bf cv} from the above input by using the single command:
%achieve the same result by using the much-shorter input file:

\begin{tcolorbox}
\footnotesize
\begin{Verbatim}[commandchars=\\\{\}]
{\bf cv}: \textcolor{green}{FERRONEMATIC_ORDER} MOLECULE_STARTS=1,3,5,7  MOLECULE_ENDS=2,4,6,8
\end{Verbatim}
\end{tcolorbox}

When PLUMED encounters this command it automatically 
%More precisely, using the short input file ensures that PLUMED 
generates the longer input file above and then uses it to perform the calculation.
This approach has three advantages:

\begin{enumerate}
\item{It reduces the amount of code that needs to be maintained}
\item{It allows us to quickly document what the FERRONEMATIC\_ORDER is computing by showing the longer version of the command that this input expands into within the PLUMED manual.} 
\item{The long version of the command is also reported in the PLUMED log file, which facilitates the identification of problems.}
\end{enumerate}

As we illustrate in the remainder of this paper, the tools described above for quickly prototyping and documenting methods allow for cross fertilization of methods, ideas and approaches across different simulation communities.  Furthermore, by reusing the same actions as much as possible we can provide the universal recommendations for optimizing performance that are the focus of the rest of this tutorial paper.

\section{How to examine PLUMED’s performance}

Before discussing our prescriptions for improving the performance of PLUMED, it is worth explaining how the measures of PLUMED’s performance that we have quoted in this tutorial have been generated.  As we mentioned in the introduction, the computational expense associated with the calculations that PLUMED is performing is often negligible when compared with the calculation of the atomic forces. Furthermore, even when the calculations PLUMED performs have a non-negligiable contribution to the total simulation time, potential non-reproducibilities in the performance
of the MD code might add noise and make PLUMED performance difficult to measure. It is thus suboptimal to run PLUMED alongside an MD code to measure its performance during the development stage.  Furthermore, although one can run stand alone analyses of trajectories using PLUMED’s driver utility, we often find that the time for such calculations is dominated by reading the trajectory.  Consequently, using the {\tt plumed~driver} command to benchmark PLUMED is also misguided.  

For these reasons, in PLUMED 2.10 we introduced a command line tool called {\tt plumed~benchmark} for reliably measuring performance across different variants of the PLUMED library.
To get the graphs of performance in this paper we have used variations on the following command when employing this tool:

\begin{Verbatim}[commandchars=\\\{\}]
plumed benchmark --plumed plumed.dat --natoms 1000 --atom-distribution sc
\end{Verbatim}

This command instructs PLUMED to repeatedly perform the calculations in the input file called {\tt plumed.dat} for a system of 1000 atoms that are arranged in an simple cubic structure.  Consequently, the same set of positions are passed to PLUMED on every step but these positions are stored in memory so there is no need to do any molecular dynamics or disk access.

{\tt plumed benchmark} has a number of features that may be useful for code developers who are worried about performance.  Please note, first and foremost, that you can read the atomic positions that should be passed to PLUMED from most of the available trajectory formats.  Consequently, if you are developing some exotic method to examine proteins or other complex molecules you can benchmark using a structure that is more relevant to the problem at hand than a simple cubic crystal.

PLUMED benchmark also allows you to run with multiple versions of PLUMED in parallel as illustrated below:

\begin{Verbatim}[commandchars=\\\{\}]
plumed-runtime benchmark --kernel /path/to/libplumedkernel.so:this 
\end{Verbatim}

When you use this command, PLUMED alternates between performing the calculations using the version of PLUMED that is in the system PATH and the version of PLUMED in {\tt /path/to/libplumedkernel.so}.  The alternation is implemented to minimize the impact of one computer's load on the relative performance of the two versions that are being compared. Once the calculation is finished timings for the calculations with the two (or more) versions of the code are output to the log.
The values PLUMED benchmark reports try to offset the initialization cost by not including it in the timings, and report an error in the relative performance of different PLUMED versions estimated using bootstrap \cite{bootstrap}.
Running such benchmark calculations to compare stable and development versions of the code is obviously useful if you are working on trying to optimise performance.

\section{Calculating multiple distances, angles, and torsions}

There are two ways to use PLUMED to calculate the three distances between atom 1 and atoms 2, 3 and 4.  You can use three actions that each pass out a single scalar as in the input below:

\begin{tcolorbox}[top=-10pt]
\singlespacing \footnotesize
\begin{Verbatim}[commandchars=\\\{\}]
{\bf d1}: \textcolor{green}{DISTANCE} ATOMS=1,2
{\bf d2}: \textcolor{green}{DISTANCE} ATOMS=1,3
{\bf d3}: \textcolor{green}{DISTANCE} ATOMS=1,4
\end{Verbatim}
\end{tcolorbox}

Or you can use one action that passes out a 3 dimensional vector as in the input below.

\begin{tcolorbox}
\footnotesize
\begin{Verbatim}[commandchars=\\\{\}]
\textcolor{blue}{\bf d}: \textcolor{green}{DISTANCE} ATOMS1=1,2 ATOMS2=1,3 ATOMS3=1,4
\end{Verbatim}
\end{tcolorbox}

Similar pairs of options are available through the {\tt ANGLE} and {\tt TORSION} commands if you are calculating multiple angles or torsions. 

\begin{figure}
    \centering
    \includegraphics[width=0.6\textwidth]{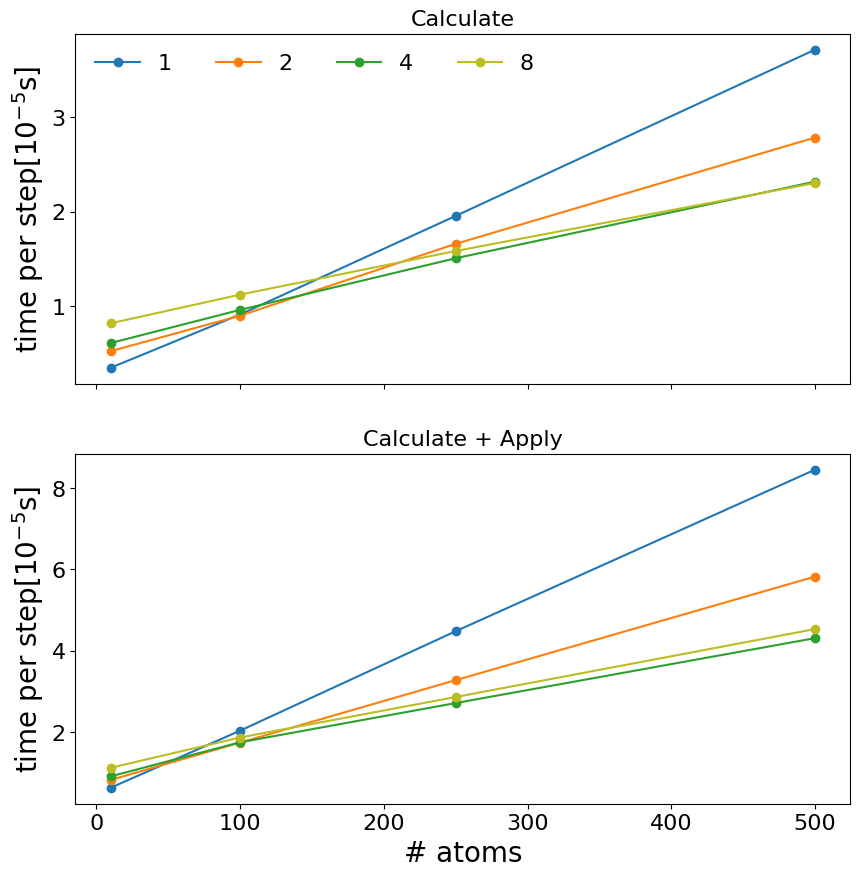}
    \caption{Time per step as a function of the number of distances that are being computed.  The blue, orange, light and dark lines indicate the cost of running the calculation with 1, 2, 4 and 8 OpenMP threads respectively. The top panel indicates the cost of calculating the distances only, while the bottom panel indicates the additional cost that comes if you apply a force on the computed distances and also need to calculate derivatives. }
    \label{fig:dist-perf1}
\end{figure}

When the number of distances being computed in these inputs is small, Fig. \ref{fig:dist-perf1} suggests that you will likely get very similar performance from these two options.  However, if you are calculating a larger number of distances, the second option will provide much better performance as the calculation of the distances in this second option can be parallelized using OpenMP and MPI. The crossovers in the top panel of Fig. \ref{fig:dist-perf1} suggest that using the second input becomes important for getting the best performance out of PLUMED once you are computing approximately 100 distances. For sizes greater than this the cost of running the calculation with multiple OpenMP threads is cheaper than running the calculation on a single thread.  
If you have forces acting upon the distances using the input that allows for multi threading becomes important to performance once you are computing 50 or more distances (Fig. \ref{fig:dist-perf1} lower panel).

The input that was used to generate the scaling plots in Fig. \ref{fig:dist-perf1} is shown below:

\begin{tcolorbox}[top=-10pt]
\singlespacing \footnotesize
\begin{Verbatim}[commandchars=\\\{\}]
\textcolor{blue}{\bf d}: \textcolor{green}{DISTANCE} ATOMS1=1,2 ATOMS2=2,3 ATOMS3=3,4 \textcolor{blue}{# etc}
{\bf s}: \textcolor{green}{SUM} ARG=\textcolor{blue}{d} PERIODIC=NO
\textcolor{green}{ RESTRAINT} ARG=s KAPPA=1 AT=0
\textcolor{green}{ PRINT} ARG=s FILE=colvar
\end{Verbatim}
\end{tcolorbox}

\begin{figure}
\centering
\includegraphics[width=\textwidth]{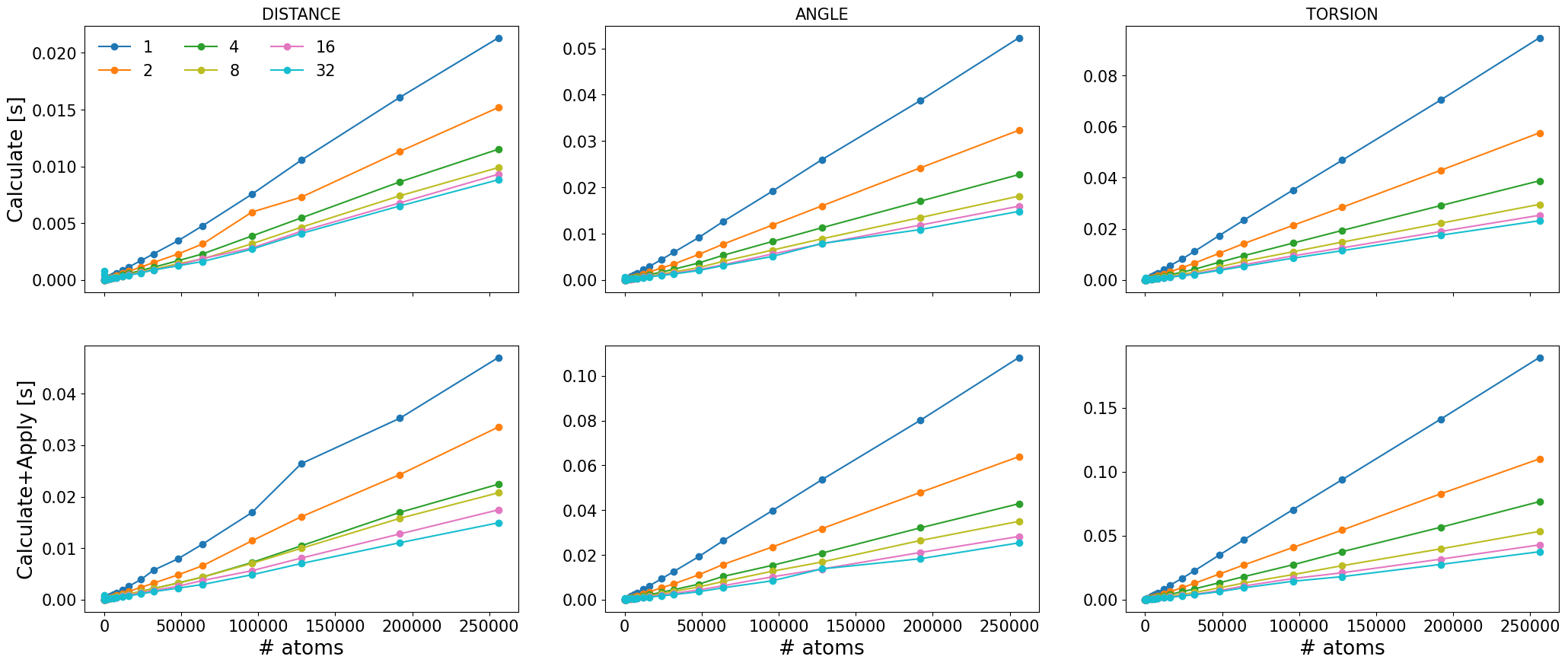}
\caption{Time taken for a single PLUMED step as a function of the number of distances (left), angles (center) and torsions (right) that are being computed.  Cost for just calculating these quantities (top panels). Cost for calculating and applying a force on the variables (bottom panels). Calculations were run on 1 - 32 OpenMP threads. The legend indicates what number of threads was used to produce each of the lines.  }
\label{fig:dist-performance}
\end{figure}

This input tells PLUMED to calculate distances for every $k$th and $(k+1)$th atom pair in the system. Consequently, if there are $n$ atoms in the system $n-1$ distances will be computed if we use the input above.  These distances are then all added together and a restraint is applied on this sum.  Similar inputs that used all $(n-2)$ sets containing the $k$th, $(k+1)$th and $(k+2)$th atoms were used to benchmark the ANGLE command, while the $(n-3)$ sets containing the $k$th, $(k+1)$th, $(k+2)$th and $(k+3)$th atoms were used to benchmark the TORSION command.  The results obtained from these calculations are shown in Fig. \ref{fig:dist-performance}      

The cost of these calculations increases linearly with the number of atoms as would be expected (Fig. \ref{fig:dist-performance}). Furthermore, having a force on these quantities roughly doubles the cost of the calculation, which, given that PLUMED recalculates all the distances/angles/torsions and their derivatives when applying forces using the chain rule, is to be expected.  Most importantly, however, for the largest systems studied you can get a roughly factor 5 speed up by using 32 OpenMP threads rather than a single thread.

\section{OpenMP versus MPI}

% Add calculations with 2 and 4 OpenMP threads to this figure to compare with MPI
% ADD pure MPI in right panel
\begin{figure}
    \centering
    \includegraphics[width=\textwidth]{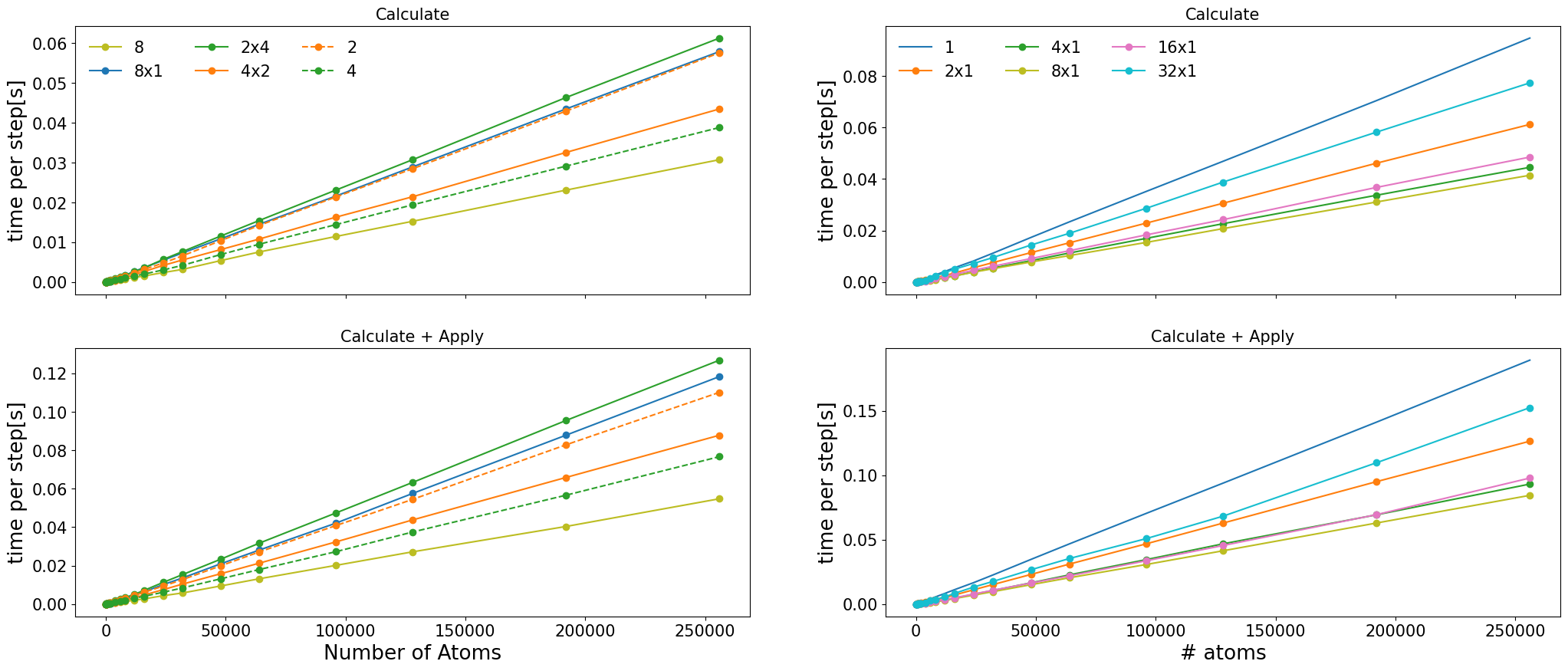}
    \caption{Time taken for a single PLUMED step as as a function of the number of torsions that are being computed.  The top panels show how the cost of calculating the torsions increases while the bottom panel shows how the cost of calculating the torsions and applying a force on these quantities changes. All the calculations that were used to generate the solid lines for graphs in the left column were run on 8 processors. For the blue line all processors communicated via OpenMP, while the orange line shows the result that was obtained when communication between the 8 processors was managed using MPI. The green and red lines show the results obtained when the two communication protocols are mixed.  The red line shows timings that are are obtained by having four MPI processors that each run on two OpenMP threads, while the green line indicates the result that is obtained by having two MPI processors running on four OpenMP threads each.  The orange and green dashed lines are results obtained when you run with 2 and 4 OpenMP threads respectively. The lines on the graphs in the right column were obtained from calculations that were parallelized over 1 to 32 MPI processes.}
    \label{fig:openmp-vs-mpi}
\end{figure}

In the previous section we noted that PLUMED calculations can be parallelized using OpenMP or MPI. 
We focused on presenting our benchmarks with different numbers of OpenMP threads rather than different numbers of MPI processes because of the results shown in the left panel of Fig. \ref{fig:openmp-vs-mpi}. To generate the lines in this figure we ran the TORSION benchmark that was introduced in the previous section using 8 OpenMP threads, 8 MPI processes, a pair of 4 OpenMP threads that communicate via MPI and four pairs of OpenMP threads that communicate via MPI. You get the best performance when you use pure OpenMP parallelism (solid light green line).  

For these relatively small calculations, the cost per step decreases when you use up to 8 MPI processors (right panel Fig. \ref{fig:openmp-vs-mpi}). However, when 16 or 32 MPI processes are used the cost of the calculation is increased by the additional communication so using fewer MPI processes is more efficient.  

We also found that calculations running over $N$ MPI processors each of which are running $M$ OpenMP threads are often slower than calculations that simply run on $M$ OpenMP threads. This is certainly the case for $N=2$ and $M=4$ but is not the case for $N=4$ and $M=2$ (dashed and solid orange and green lines left panel Fig. \ref{fig:openmp-vs-mpi}).
%The dashed lines show the cost associated with running the calculation on 2 and 4 OpenMP threads. They should be compared with the solid lines of the same colors, which show the cost of running the same calculation with 4 or 2 MPI processes that are each running on 2 or 4 OpenMP threads.  
%You can see that running two MPI processes that each run 4 OpenMP threads is actually less efficient than simply running on 4 threads. 
Consequently, if your MD code is running using a combination of OpenMP and MPI it may be worth using the SERIAL flag in the PLUMED input to turn on off all MPI parallelism in PLUMED. Having completely separate instances of the PLUMED calculations on each of the MPI processes is often faster than dividing the calculations between the MPI processes and then communicating the data to all nodes.

%Notice finally that you can also use a single line command like the one we have used above to calculate vectors of distances, to calculate vectors of angles, torsions, dipoles or vectors that define the planes containing sets of three atoms.  The performance of the single action version of all these commands will be better than the performance of the multiple action version for the same reasons that the single action version of the distance command is faster.  You can thus use the ideas discussed in this section when optimising the performance of these calculations.

\section{Working with symmetry functions} 
\label{sec:symfunc}

When studying phenomena such as crystal nucleation and growth using symmetry functions is commonplace \cite{uberti,bingqing1,bingqing2,bingqing3}.  These functions have the general form:

\begin{equation}
s_i = \frac{1}{\sum_{j=1}^N \sigma(r_{ij})} \sum_{j=1}^N f(x_{ij},y_{ij},z_{ij})\sigma(r_{ij})
\label{eqn:symfunc}
\end{equation}
where $(x_{ij}, y_{ij},z_{ij})$ is the vector connecting atoms $i$ and $j$, $r_{ij}$ is this vector's modulus and 
$\sigma$ is a continuous switching function that is one when its argument is small and 0 when its argument is large.  An example input that illustrates how such a function can be calculated using PLUMED is shown below:

\begin{tcolorbox}[top=-10pt]
\singlespacing \footnotesize
\begin{Verbatim}[commandchars=\\\{\}]
\textcolor{blue}{ # Calculate four NxN matrix called cmap.w, cmap.x, cmap.y and cmap.z}
\textcolor{blue}{ # Element ij of the matrix cmap.w is equal to sigma(r_ij)}
\textcolor{blue}{ # Element ij of the matrices cmap.x, cmap.y and cmap.z are} 
\textcolor{blue}{ # equal to x_ij, y_ij and z_ij respectively. }
\textcolor{red}{\bf cmap}: \textcolor{green}{CONTACT_MATRIX} ...
  GROUP=@mdatoms SWITCH=\{RATIONAL D_0=0.6 R_0=1 NN=6 MM=12 D_MAX=2.0\} 
  COMPONENTS
...
\textcolor{blue}{ # Calculate a matrix r whose ij element is equal to r_ij}
\textcolor{red}{\bf r}: \textcolor{green}{CUSTOM} ...
   ARG=\textcolor{red}{cmap.x},\textcolor{red}{cmap.y},\textcolor{red}{cmap.z} 
   FUNC=sqrt(x*x+y*y+z*z)
   PERIODIC=NO
...
\textcolor{blue}{ # Calculate a matrix called f whose ij element is equal to the } 
\textcolor{blue}{ # quantity inside the sum of the numerator above.} 
\textcolor{red}{\bf f}: \textcolor{green}{CUSTOM} ...
    ARG=\textcolor{red}{cmap.w},\textcolor{red}{cmap.x},\textcolor{red}{cmap.y},\textcolor{red}{cmap.z},\textcolor{red}{r} VAR=w,x,y,z,r
    FUNC=w*((x^4+y^4+z^4)/(r^4))
    PERIODIC=NO
...
\textcolor{blue}{ # Evaluate the sum in the numerator expression above by multiplying the matrix}
\textcolor{blue}{ # that is computed by the above action by a vector of ones.}
\textcolor{blue}{\bf ones}: \textcolor{green}{ONES} SIZE=@natoms
\textcolor{blue}{\bf numer}: \textcolor{green}{MATRIX_VECTOR_PRODUCT} ARG=\textcolor{red}{f},\textcolor{blue}{ones}
\textcolor{blue}{ # Evaluate the denominator in the expression above in a similar way}
\textcolor{blue}{\bf denom}: \textcolor{green}{MATRIX_VECTOR_PRODUCT} ARG=\textcolor{red}{cmap.w},\textcolor{blue}{ones}
\textcolor{blue}{ # And finally evaluate the values of the order parameter above}
\textcolor{blue}{ # for each of the 64 atoms}
\textcolor{blue}{\bf op}: \textcolor{green}{CUSTOM} ARG=\textcolor{blue}{numer},\textcolor{blue}{denom} FUNC=x/y PERIODIC=NO
\textcolor{blue}{ # Now calculate the mean of all the order parameters}
{\bf mean}: \textcolor{green}{MEAN} ARG=\textcolor{blue}{op} PERIODIC=NO
\textcolor{blue}{ # And add a bias}
\textcolor{green}{BIASVALUE} ARG=mean
\end{Verbatim}
\end{tcolorbox}
    
Notice that in inputs such as the one above we are now passing matrices between actions as well as scalars and vectors. We use red to distinguish the labels of matrices from the vectors and scalars.  Further note, that in the same way as we did for the {\tt FERRONEMATIC\_ORDER} parameter that was discussed in section 2, we provide shortcuts that hide this complex input from casual users.  Importantly, however, decomposing the calculation in the manner shown in the above input allows us to use the same or similar actions for many different symmetry functions. Furthermore, by optimizing these common actions we improve performance for many different symmetry function types.

The first and most important of these tricks is the use of the {\tt D\_MAX} parameter in the input to the switching function that is used in the {\tt CONTACT\_MATRIX} command.  A {\tt D\_MAX} value can be set whenever you define a switching function in PLUMED. By setting this parameter of the switching function you are enforcing the value of the switching function to be zero for all $r>d_{max}$ by using the stretching and scaling function that is computed from the switching function, $\theta$, as follows.

\begin{equation}
s(r_{ij}) = \frac{\theta(r_{ij}) - \theta(d_{max})}{\theta(0) - \theta(d_{max})}
\label{eqn:sij}
\end{equation}

Consequently, when we evaluate $\sigma(r_ij)$ in the expression above we are not computing the usual rational switching function:

\begin{equation}
\theta(r_{ij}) = \frac{1}{1 + \left(\frac{r_{ij}}{r_0}\right)^{6}}
\label{eqn:thetaij}
\end{equation}

$\sigma(r_{ij})$ is instead the product of the $s(r_{ij})$ and $\theta(r_{ij})$ functions that were defined in equations \ref{eqn:sij} and \ref{eqn:thetaij}.

\begin{figure}
\centering
\includegraphics[width=0.6\textwidth]{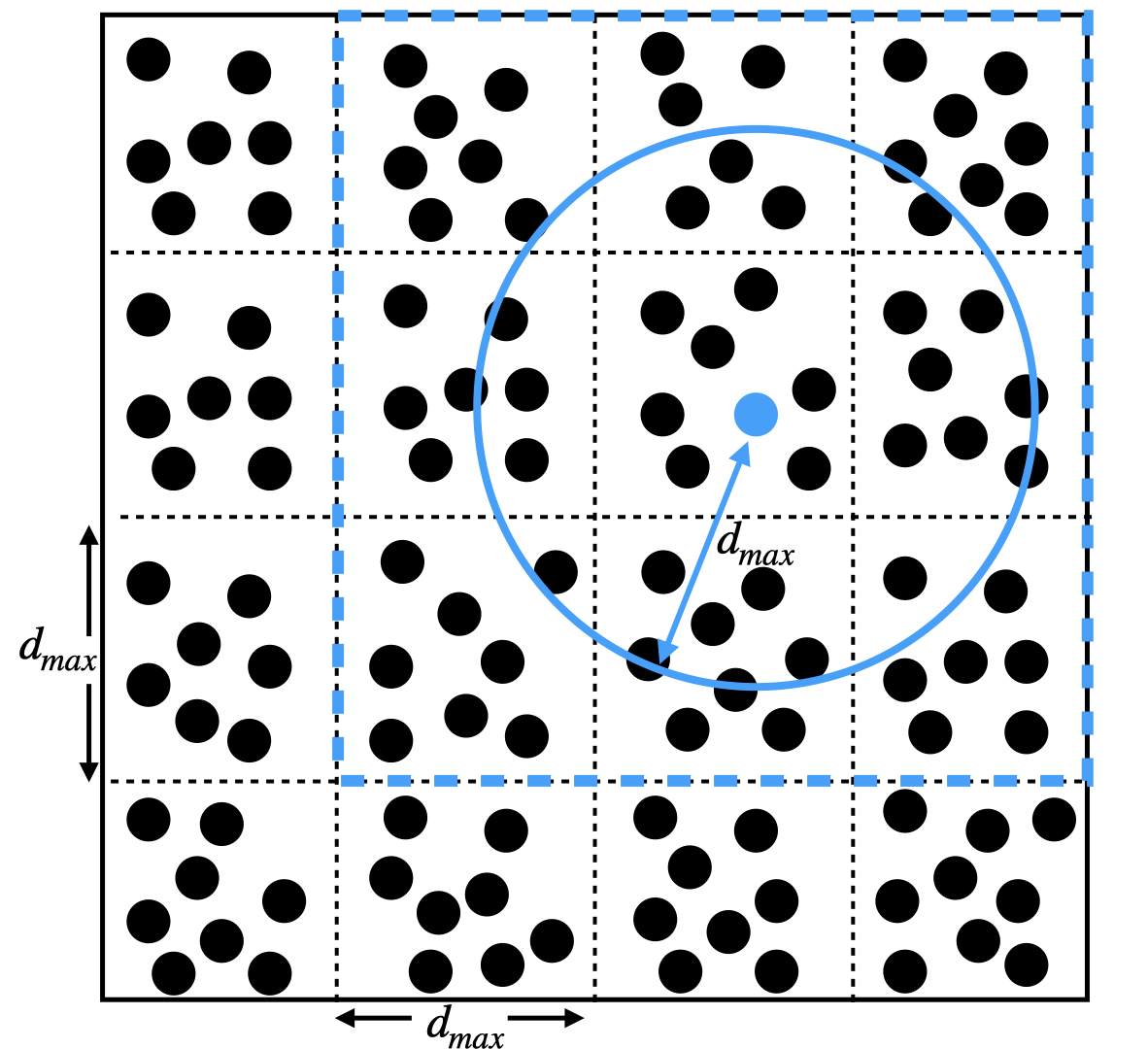}
\caption{The linked list algorithm that is employed within PLUMED. The cell box is divided into a set of cubes with a side length of $d_{max}$. The cell each atom resides within is then determined at the start of the calculation. This reduces computational expense when we compute contact matrices because we can determine the neighbors of the blue atom by iterating over the atoms in the cell the blue atom resides in and the atoms in this cell's immediate neighbors. The distance between the blue atom and any atom that is not in the same cell or one of its neighbors is guaranteed to be greater than $d_{max}$. }
\label{fig:linked-list}
\end{figure}

The fact that $\sigma(r_{ij})$ is guaranteed to be zero for all $r_{ij}>d_{max}$ ensures that we can use the linked list strategy that is illustrated in Fig. \ref{fig:linked-list} to optimize the calculation of the contact matrix.  This strategy works by first dividing the simulation cell into cubic boxes that each have a side length of $d_{max}$. The box each of the input atoms resides in is then identified.  When we evaluate the $i$th row of the contact matrix we only evaluate element $i,j$ if atom $j$ is in the same box as atom $i$ or one of the 26 boxes that are adjacent to the box that contains atom $i$.  When {\tt D\_MAX} is used the calculation of the symmetry functions thus scales linearly and not quadratically with the number of atoms.  Furthermore, because we are normally only interested in the structure in the first coordination sphere around the atoms, the {\tt D\_MAX} value can be set to a relatively small value. For the example calculations in this tutorial, which are run on an ideal simple cubic structure with a lattice parameter of 1 nm, {\tt D\_MAX} is set equal to 2~nm so the sum in equation \ref{eqn:symfunc} runs over the 18 atoms that are 1, $\sqrt{2}$ or $\sqrt{3}$~nm from the central atom. The boxes for our linked list algorithm are thus considerably smaller than those one would be using when exploiting similar tricks for evaluating the interatomic forces in an MD code. 

\begin{figure}
\centering
\includegraphics[width=0.6\textwidth]{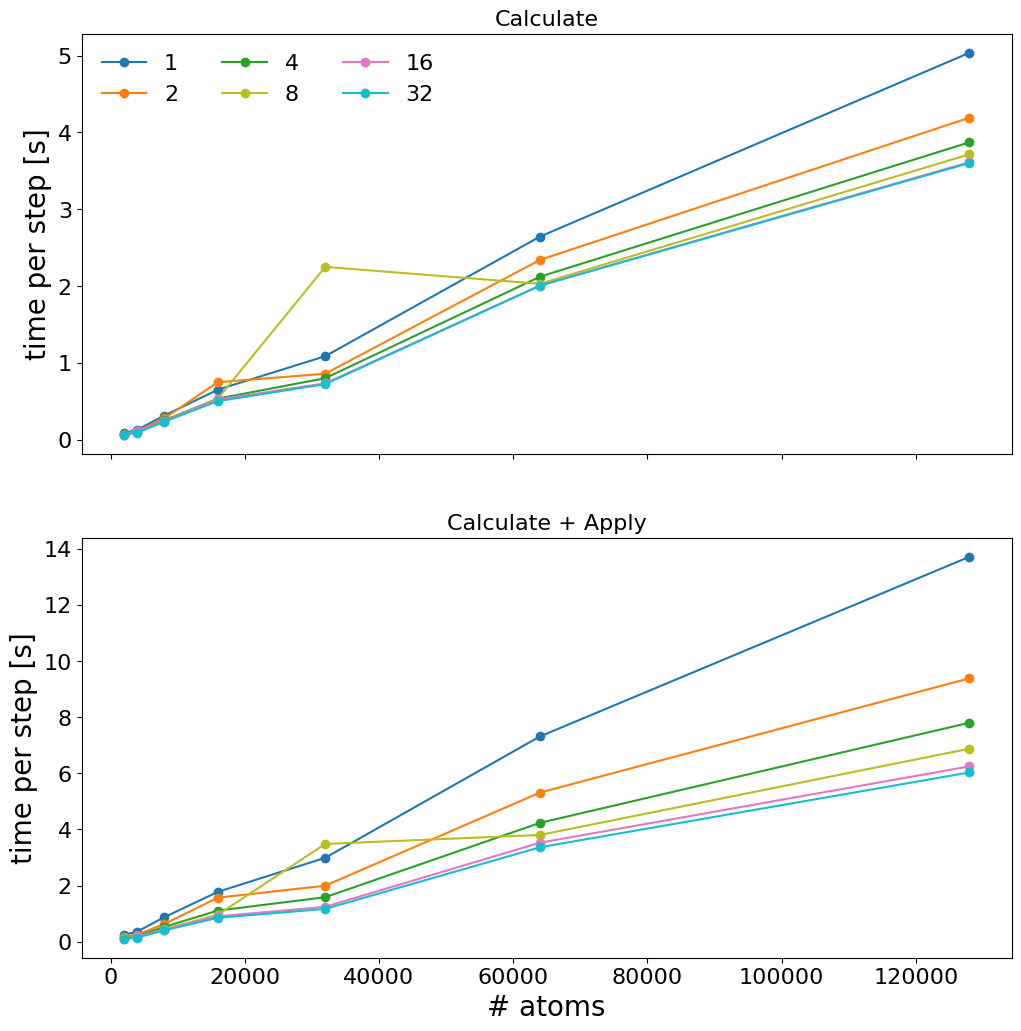}
\caption{Graphs indicating how the cost of a single PLUMED step changes as you increase the number of symmetry functions that are being computed.  The top panel shows how the cost of calculating the symmetry functions changes, while the bottom panel shows how the cost of calculating the symmetry function and applying a force upon it changes. The various lines show the costs when the calculation is run on the numbers of OpenMP threads indicated in the legend. }
\label{fig:symfunc-performance}
\end{figure}

Using {\tt D\_MAX} in the way described above also ensures that we can use sparse matrix storage for the \textcolor{red}{\bf cmap.w}, \textcolor{red}{\bf cmap.x}, \textcolor{red}{\bf cmap.y}, \textcolor{red}{\bf cmap.z}, \textcolor{red}{\bf r} and \textcolor{red}{\bf f} matrices in the above input and sparse matrix algebra when applying a functions to the elements of the matrix in the CUSTOM actions and when multiplying these matrices by vectors in the {\tt MATRIX\_VECTOR\_PRODUCT} actions to further improve performance.

When using an input such as the one above you can parallelize the calculation of all actions using OpenMP and MPI. Simulations were performed to determine how the time it takes PLUMED to perform a calculation using 1-32 OpenMP threads with the input above depends on the number of atoms that are input to the {\tt CONTACT\_MATRIX} command (Fig. \ref{fig:symfunc-performance}).  You can see that adding a force on the symmetry functions increases the cost of the calculation by roughly a factor of 3.  However, for the largest systems, using OpenMP reduces the cost of the calculation of the forces by a factor of 3. Even so, the cost of this calculation, when run with the large numbers of atoms that have been used here, is likely too great for it to be used as a CV in an MD simulation. However, the implementation discussed here and the implementations of other expensive quantities discussed in this paper can be used to generate training data for a cheaper-to-evaluate neural network as discussed in \cite{matteo-machine-learning}.  

\section{Evaluating symmetry functions in a particular part of the box using the MASK keyword}
\label{sec:volumes}

To reduce the computational expense associated with the calculation of symmetry functions some developers sometimes choose to only evaluate the values of symmetry functions for those atoms in a particular part of the box \cite{uberti}. This approach makes particular sense if one is examining nucleation at a surface \cite{heteroNucleation} or if one is using a restraint to prevent a seed from dissolving \cite{seedMD}.  This approach would also be necessary if one were computing symmetry functions when using the methods for running at constant chemical potential discussed in \cite{constantChemicalPotential}.

The problem when using such approaches is that the atoms within the region of interest, for which the symmetry function has to be evaluated, changes dynamically as the simulation progresses and atoms exchange in and out of the region of interest.  We consequently need some way for dynamically indicating the set of atoms for which the symmetry functions need to be evaluated.  The following example input illustrates how the MASK keyword can be used for precisely this purpose:

\begin{tcolorbox}[top=-10pt]
\singlespacing \footnotesize
\begin{Verbatim}[commandchars=\\\{\}]
\textcolor{blue}{\bf ones}: \textcolor{green}{ONES} SIZE=@natoms
\textcolor{blue}{ # Create an atom that is fixed at the origin}
\textcolor{violet}{\bf center}: \textcolor{green}{FIXEDATOM} AT=0,0,0
\textcolor{blue}{ # Determine if each of the atoms is within a sphere of radius 1.5 nm that is}
\textcolor{blue}{ # centered on the point (0,0,0)}
\textcolor{blue}{\bf w}: \textcolor{green}{INSPHERE} ...
    ATOMS=@mdatoms CENTER=\textcolor{violet}{center} 
    RADIUS=\{RATIONAL D_0=24.9 R_0=0.01 D_MAX=25\}
...
\textcolor{blue}{ # Now evaluate the order parameters }
\textcolor{red}{\bf cmap}: \textcolor{green}{CONTACT_MATRIX} ...
  GROUP=@mdatoms SWITCH=\{RATIONAL D_0=0.6 R_0=1 NN=6 MM=12 D_MAX=2.0\} 
  COMPONENTS MASK=\textcolor{blue}{w}
...
\textcolor{red}{\bf r}: \textcolor{green}{CUSTOM} ...
   ARG=\textcolor{red}{cmap.x},\textcolor{red}{cmap.y},\textcolor{red}{cmap.z} 
   FUNC=sqrt(x*x+y*y+z*z)
   PERIODIC=NO
...
\textcolor{red}{\bf f}: \textcolor{green}{CUSTOM} ...
    ARG=\textcolor{red}{cmap.w},\textcolor{red}{cmap.x},\textcolor{red}{cmap.y},\textcolor{red}{cmap.z},\textcolor{red}{r}
    VAR=w,x,y,z,r
    FUNC=w*((x^4+y^4+z^4)/(r^4))
    PERIODIC=NO
...
\textcolor{blue}{\bf numer}: \textcolor{green}{MATRIX_VECTOR_PRODUCT} ARG=\textcolor{red}{f},\textcolor{blue}{ones}
\textcolor{blue}{\bf denom}: \textcolor{green}{MATRIX_VECTOR_PRODUCT} ARG=\textcolor{red}{cmap.w},\textcolor{blue}{ones}
\textcolor{blue}{ # Evaluate the order parameter multiplied by the vector of ones and zeros}
\textcolor{blue}{ # that tells you whether or not each atom is in the region of interest}
\textcolor{blue}{\bf op}: \textcolor{green}{CUSTOM} ARG=\textcolor{blue}{w},\textcolor{blue}{numer},\textcolor{blue}{denom} FUNC=x*(y/z) PERIODIC=NO
{\bf opsum}: \textcolor{green}{SUM} ARG=\textcolor{blue}{op} PERIODIC=NO
{\bf vsum}: \textcolor{green}{SUM} ARG=\textcolor{blue}{w} PERIODIC=NO
\textcolor{blue}{ # Evaluate the average value of the order parameter for those atoms that} 
\textcolor{blue}{ # are in the region of interest.}
{\bf mean}: \textcolor{green}{CUSTOM} ARG=opsum,vsum FUNC=x/y PERIODIC=NO
\textcolor{green}{BIASVALUE} ARG=mean  
\end{Verbatim}
\end{tcolorbox}

The general form for the order parameter that is being evaluated here is given by the following expression:

$$
\xi = \frac{\sum_i w(x_i) s_i}{\sum_i w(x_i)}.
$$

In this expression $s_i$ is the symmetry function that is defined in equation \ref{eqn:symfunc}.  $w(x_i)$ is then a differentiable function of the position, $x_i$, of atom $i$ that is one if the atom is in the region of interest and zero otherwise.  In the input above this $w(x_i)$ function is a switching function that acts upon the distance between atom $i$ and the origin of the lab frame.  The $i$th element of the vector, \textcolor{blue}{\bf w}, is thus one if it is within a sphere centered on the origin and zero otherwise.

The important thing to note in this input is that the vector \textcolor{blue}{\bf w} is passed to the {\tt CONTACT\_MATRIX} action through the {\tt MASK} keyword even though it is not needed to calculate the contact matrix.  Passing this vector in this way ensures that the {\tt CONTACT\_MATRIX} only calculates the $i$th row of the matrix if the $i$th element in \textcolor{blue}{\bf w} is non zero.  In other words, by passing the vector \textcolor{blue}{\bf w} through the {\tt MASK} keyword we ensure that $s_i$ values are only computed for those atoms that are in the sphere of interest.

\begin{figure}
\includegraphics[width=0.8\textwidth]{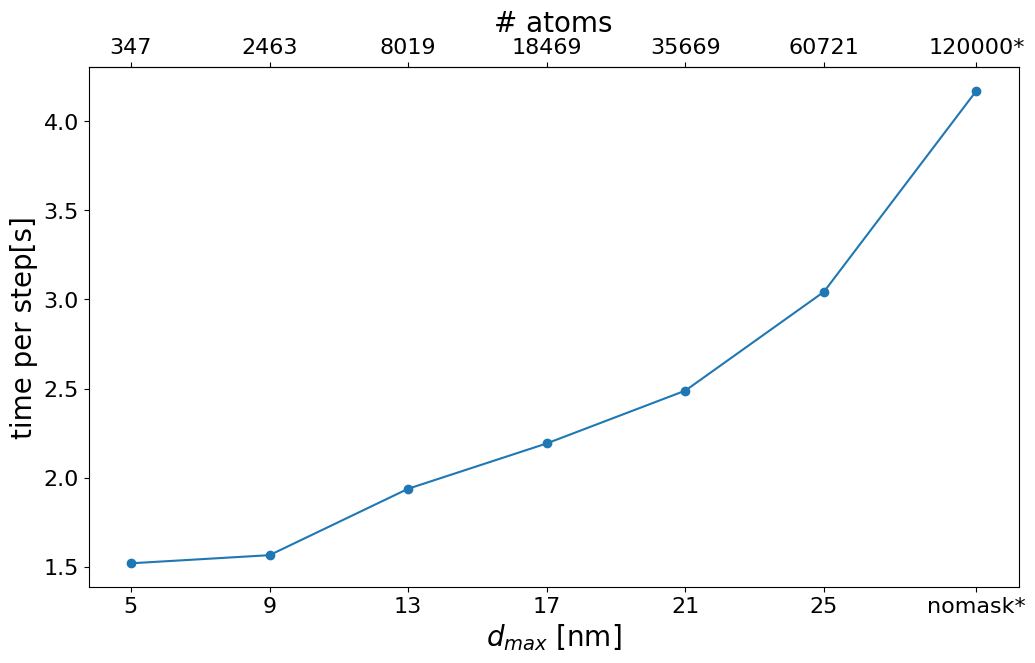}
\caption{The cost of a single PLUMED step changes as a function of the volume of the spherical region in which you are evaluating symmetry functions for the atoms.  The bottom $x$-axis indicates the radius of the spherical region in which the symmetry functions are being evaluated, while the top $x$ axis indicates the number of atoms for which symmetry functions are being evaluated.}
\label{fig:mask-performance}
\end{figure}

To determine the effect this trick has on computational performance we used PLUMED running on 16 OpenMP threads to calculate the average value of the symmetry function in a spherical sub-region of a system of 120000 atoms (Fig. \ref{fig:mask-performance}). The bottom $x$-axis in this figure indicates the radius of the sphere in which the symmetry function is being evaluated, while the top axis is then used indicate the number of atoms that are within a sphere of this radius. You can clearly see how the cost of the calculation is reduced as the radius of the spherical region of interest decreases and the number of atoms for which equation \ref{eqn:symfunc} is being evaluated decreases.   

\section{Another use for the MASK keyword}

The example provided in the previous section illustrates one application of a common approach for working with vectors and matrices whose elements are a product of a part that is cheap to evaluate and a part that is more expensive to evaluate. As illustrated above, if you are working with such objects you first evaluate the object with the computationally cheap elements and determine if any of the elements of this vector are zero. You then use the value from this cheap action as a mask that tells the more computationally expensive action that there are elements of its output that do not need to be computed.

Another place where this approach is used in PLUMED is in the implementation of the STRANDS\_CUTOFF keyword in the {\tt ANTIBETARMSD} and {\tt PARABETARMSD} actions \cite{secondarystructure}.  The following example is a typical input that uses this keyword:

\begin{tcolorbox}[top=-10pt]
\singlespacing \footnotesize
\begin{Verbatim}[commandchars=\\\{\}]
\textcolor{green}{MOLINFO} STRUCTURE=protein.pdb
{\bf ab:} \textcolor{green}{ANTIBETARMSD} ...
   RESIDUES=all TYPE=OPTIMAL
   STRANDS_CUTOFF=1.0 R_0=0.1 NN=8 MM=12
...
\textcolor{green}{BIASVALUE} ARG=ab
\end{Verbatim}
\end{tcolorbox}

The {\tt ANTIBETARMSD} command that is used here is an example of a shortcut action.  When PLUMED reads this input it converts it into the input for a set of actions that together compute the {\tt ANTIBETARMSD} collective variable which is defined as follows:

\begin{equation}
s = \sum_i \frac{1 - \left(\frac{R(\mathbf{X}_i,\mathbf{X}_{ref})}{r_0}\right)^8}{1 - \left(\frac{R(\mathbf{X}_i,\mathbf{X}_{ref})}{r_0}\right)^{12}}
\label{eqn:secondarystructure}
\end{equation}

In this expression the sum runs over all the six residue segments of protein that could conceivably form an antiparallel beta sheet and $\mathbf{X}_i$ is the positions of the 30 backbone atoms in each of these residue segments. $\mathbf{X}_{ref}$ is the positions of the 30 backbone atoms in an ideal antiparallel beta sheet so $R(\mathbf{X}_i,\mathbf{X}_{ref})$ is the RMSD distance between the instantaneous configuration of the backbone atoms in the $i$th residue segment and the ideal structure for an antiparallel beta sheet. The sum in equation \ref{eqn:secondarystructure} thus counts how many segments of the protein resemble an antiparallel beta sheet.  

\begin{figure}
\centering
\includegraphics[width=\textwidth]{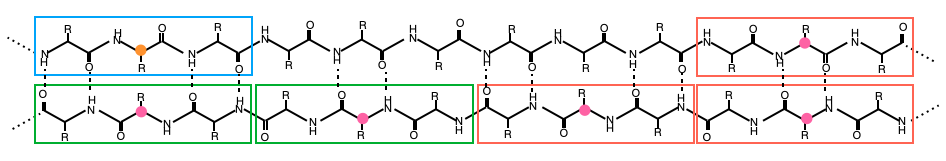}
\caption{The method via which the {\tt STRANDS\_CUTOFF} keyword of {\tt ANTIBETARMSD} improves the performance of this action. PLUMED needs to determine if the three residue segment in the blue rectangle forms an antiparallel beta sheet with each of the three residues in each of the red and green rectangles by computing $R(\mathbf{X}_i,\mathbf{X}_{ref})$. However, before computing these $R(\mathbf{X}_i,\mathbf{X}_{ref})$ values, PLUMED computes the distance between the yellow atom and each of the pink atoms. The value of $R(\mathbf{X}_i,\mathbf{X}_{ref})$ is then only computed if this distance is less than the {\tt STRANDS\_CUTOFF} value. Consequently, we compute two $R(\mathbf{X}_i,\mathbf{X}_{ref})$ values rather than five values as the central atom in the three-residue segments that are in red rectangles are too far from the three-residue segment in the blue rectangle to possibly form an antiparallel beta sheet.}
\label{fig:antibetermsd}
\end{figure}

PLUMED assumes that every pair of three residue segments that are separated by more than six residues along the chain can form an antiparallel beta sheets (Fig. \ref{fig:antibetermsd}).  This action is thus computationally expensive because number of potential antiparallel beta sheets scales quadratically with the number of residues. 

\begin{figure}
\includegraphics[width=0.3\textwidth]{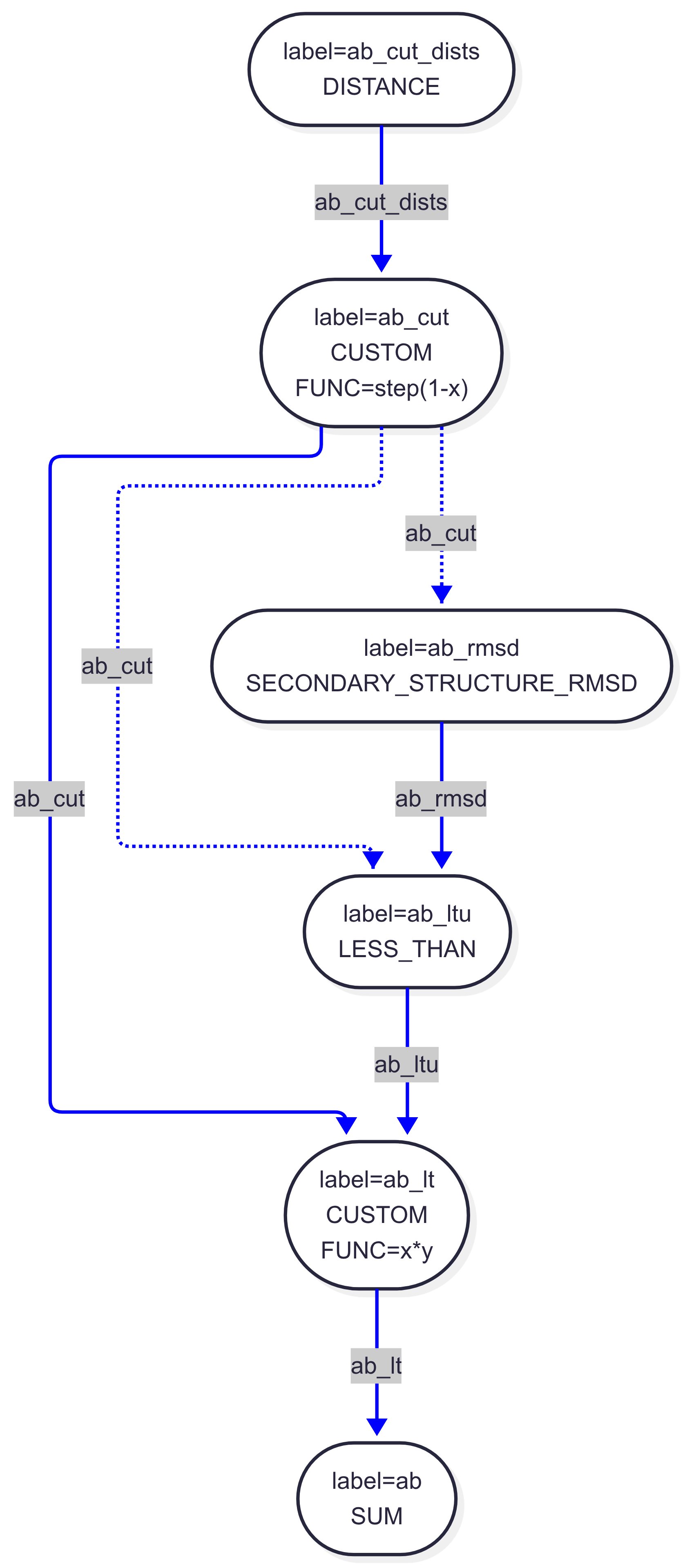}
\caption{The actions used to evaluate secondary structure variables.}
\label{fig:ssrmsd-graph}
\end{figure}

The particular set of actions that are used to compute the {\tt ANTIBETARMSD} collective variable function and the way the values are passed between them are shown in Fig. \ref{fig:ssrmsd-graph}.  Notice the dashed line that connects the {\tt CUSTOM} action with label \textcolor{blue}{\bf ab\_cut} and the {\tt SECONDARY\_STRUCTURE\_RMSD} action with label \textcolor{blue}{\bf ab\_rmsd}.  This line illustrates that the vector \textcolor{blue}{\bf ab\_cut} is being used as a MASK in the second action.  Each element in this vector is only equal to one if the distance between the C alpha atoms of the two central residues of the three-residue segments that we are comparing to an idealized antiparallel beta sheet is less than a cutoff. If this distance is larger than the cutoff then the element is zero.  Consequently, by using this vector as a mask on the {\tt SECONDARY\_STRUCTURE\_RMSD} action we ensure that the expensive calculation of $R(\mathbf{X}_i,\mathbf{X}_{ref})$ in equation \ref{eqn:secondarystructure} above is only performed for a small subset of the residues in the protein, which could potentially form an antiparallel beta sheet. 

%In other words, to optimise this calculation we do not calculate the RMSD values if the two protein strands are very apart and the structure is thus very different from an anitparallel beta sheet.
 
To understand why this works consider the atoms involved in five of the $R(\mathbf{X}_i,\mathbf{X}_{ref})$ values that we would have to calculate to obtain {\tt ANTIBETARMSD} without {\tt STRANDS\_CUTOFF} that are shown in Fig. \ref{fig:antibetermsd}. In evaluating equation \ref{eqn:secondarystructure} we need to consider whether the atoms in the blue rectangle form an antiparallel beta sheet with each of the three-residue segments shown in the green and red rectangles.  However, if we compute the distances between the yellow atom and each of the pink atoms we immediately see that the configurations formed by the the residues in the blue rectangle and each of the red rectangles cannot possibly resemble an antiparallel beta sheet as the central atoms in the residues are far too far apart.  To compute equation \ref{eqn:secondarystructure} we thus only need to compute the two $R(\mathbf{X}_i,\mathbf{X}_{ref})$ values that involve the atoms in the blue rectangle and the atoms in the two green rectangles.       

\begin{figure}
\centering
\includegraphics[width=0.7\textwidth]{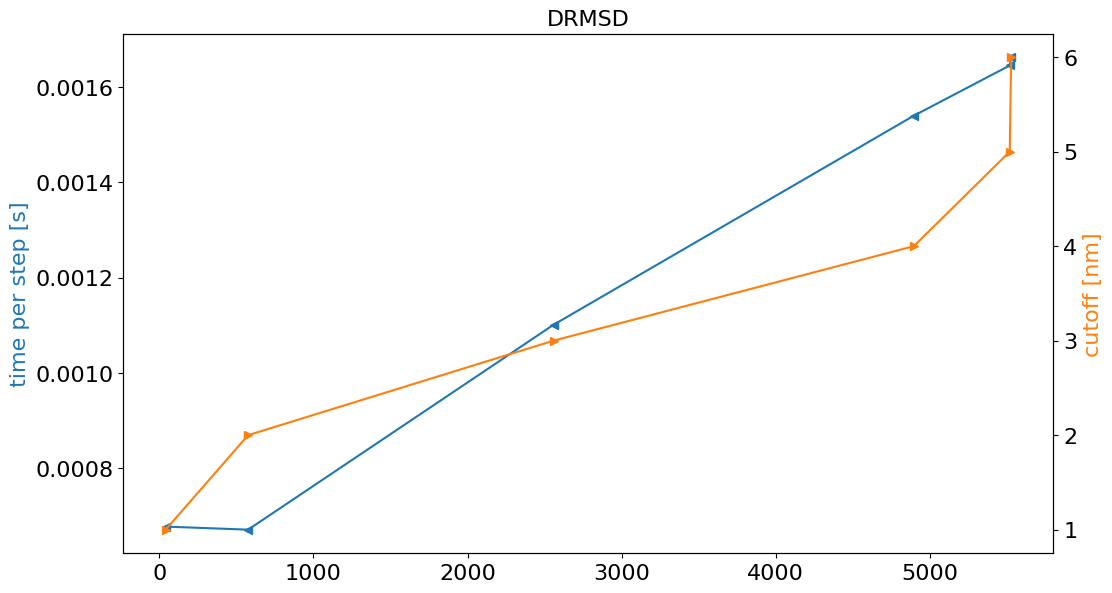}
\caption{Lowering the {\tt STRANDS\_CUTOFF} value reduces the computational cost of the {\tt ANTIBETARMSD} action. The orange line and right axis give the values we used for this cutoff. As discussed in the text, when you use a smaller {\tt STRANDS\_CUTOFF} you need to do fewer expensive RMSD or DRMSD calculations.  The $x$-axis indicates how many of these DRMSD calculations are being performed for each cutoff.  The blue line then shows how the time to perform these calculations changed as we increased the {\tt STRANDS\_CUTOFF} value.  You can see that reducing this quantity to a reasonable value of 1.0~nm reduces the cost of the calculation by more than a factor of two as you move from needing to calculate over 5000 D/RMSD values to having to calculate less than 1000.}
\label{fig:secondarystructure-performance}
\end{figure}
 
By using the STRANDS\_CUTOFF keyword correctly we can improve the performance of the {\tt ANTIBETARMSD} action by a factor of two for a small protein system with only 180 residues (Fig. \ref{fig:secondarystructure-performance}).  We have plotted the performance of the version of this CV that was implemented in the paper where Pietrucci and Laio \cite{secondarystructure} first introduced this variable which used the DRMSD to measure the distances between the instantaneous and reference structure in figure \ref{fig:secondarystructure-performance}. A revised version of this CV that we implemented in PLUMED, that uses the RMSD in place of the DRMSD is also available within PLUMED. The RMSD version of this CV is slightly cheaper than the DRMSD version but the difference in cost is marginal.        

\section{Steinhardt parameters}

Steinhardt parameters \cite{steinhardt,delago_steinhardt,clusterpaper} are a key component of many approaches for studying nucleation in atomic systems.  This approach is often claimed to be superior to the approach based on the symmetry function that was introduced earlier because it accounts for rotational invariance.  These rotational invariances are accounted for by computing all $(2l+1)$ spherical harmonics, $Y_l^m$, for a particular $l$ value using: 
$$
q_{lm}(i) = \sum_{j=1}^N \sigma(r_{ij}) Y_l^m(\theta_{ij},\phi_{ij}) \quad \textrm{where} \quad \theta_{ij} = \cos^{-1}\left( \frac{z_{ij}}{r_{ij}}\right) \quad \textrm{and} \quad e^{i\phi_{ij}} = \frac{x_{ij}}{r_{ij}} + i \frac{y_{ij}}{r_{ij}}.
$$
Notice that this equation resembles equation \ref{eqn:symfunc}, which was introduced in the section on symmetry functions.  In writing it we have thus used the symbols that were defined when we introduced that earlier equation.

From these $(2l+1)$ complex numbers one can then compute a modulus using: 
\begin{equation}
Q_l(i) = \frac{|q_l(i)|}{\sum_j \sigma(r_{ij})} \quad \textrm{where} \quad  |q_l(i)| = \sqrt{\sum_{m=-l}^l q_{lm}(i)^* q_{lm}(i)}
\label{eqn:q6}
\end{equation}
for each of the individual atoms.  Alternatively, one can compute the following product of the $(2l+1)$ spherical harmonics on atoms $i$ and $j$ \cite{tenWolde}.

\begin{equation}
Q(i,j) = \sum_{m=-l}^l \left(\frac{q_{lm}(i)}{|q_l(i)|} \right)^* \left(\frac{q_{lm}(j)}{|q_l(j)|} \right)
\label{eqn:lq6}
\end{equation}

The advantage of this second approach over the first is that a comparison of the orientations of the coordination spheres around atoms $i$ and $j$ is performed.  In other words, the dot product that is evaluated in the expression above is only large if the same $q_{lm}(i)$ and $q_{lm}(j)$ values are large on both atom $i$ and atom $j$.  This second approach is thus less affected if there are a significant number of $q_{lm}(i)$ components with moderately large values than the first.

The example input below illustrates how we can use PLUMED to calculate the average $Q_6(i)$ value for all the atoms in the system.

\begin{tcolorbox}[top=-10pt]
\singlespacing \footnotesize
\begin{Verbatim}[commandchars=\\\{\}]
\textcolor{blue}{\bf q6}: \textcolor{green}{Q6} SPECIES=@mdatoms SWITCH=\{RATIONAL D_0=0.6 R_0=1 NN=6 MM=12 D_MAX=2.0\}
{\bf mean}: \textcolor{green}{MEAN} ARG=\textcolor{blue}{q6} PERIODIC=NO
 \textcolor{green}{BIASVALUE} ARG=mean
\end{Verbatim}
\end{tcolorbox}

\begin{figure}
\includegraphics[width=0.6\textwidth]{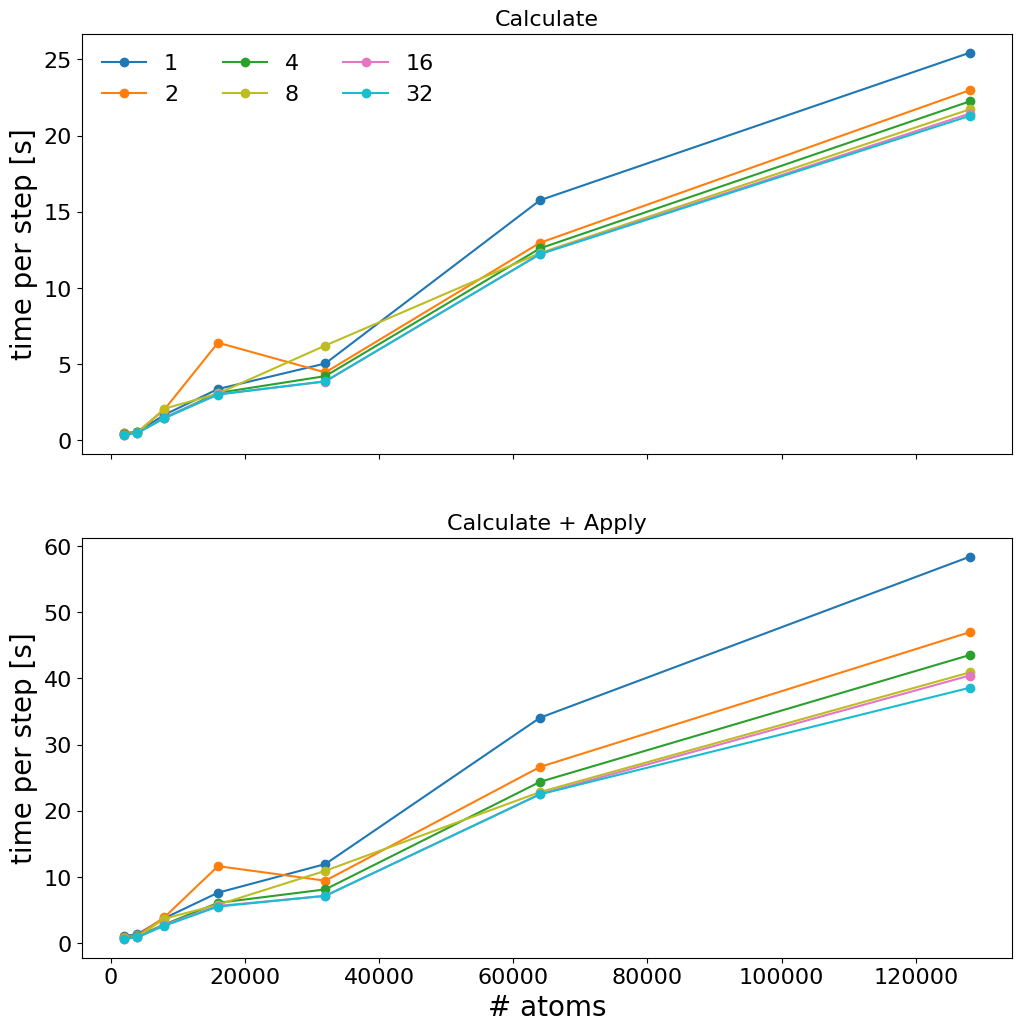}
\caption{The cost of a single PLUMED step as a function of the number of $Q_6$ parameters that are being computed. The top panel shows how the cost of calculating the $Q_6$ parameters changes, while the bottom panel shows how the cost of calculating $Q_6$ and applying a force upon it changes. The various lines show the costs when the calculation is run on the numbers of OpenMP threads indicated in the legend.}
\label{fig:q6-performance}
\end{figure}

The input here is a shortcut once again.  An expanded version that does something similar to this shortcut is provided in the next example input below. We ran calculations to determine how the computational expense associated with evaluating this variable changes with system size (Fig. \ref{fig:q6-performance}. A comparison of Fig. \ref{fig:symfunc-performance} and \ref{fig:q6-performance} illustrates that computing $Q_6$ is roughly six times more expensive than computing a symmetry function. 

As discussed above, order parameters that use equation \ref{eqn:lq6} in place of equation \ref{eqn:q6} usually demonstrate superior performance at distinguishing between atoms that are part of the solid and liquid phases. A typical order parameter for an atom that is computed using equation is given by:

\begin{equation}
s_i = \frac{\sum_{j=1}^n \sigma(r_{ij})Q(i,j)}{\sum_{j=1}^n \sigma(r_{ij})} 
\label{eqn:lq6-proper}
\end{equation}

To compute the average value of this order parameter for all 1000 of the atoms in the system using PLUMED we would use the following input.

\begin{tcolorbox}[top=-10pt]
\singlespacing \footnotesize
\begin{Verbatim}[commandchars=\\\{\}]
\textcolor{blue}{ # Compute sigma(r_ij) and the (x_ij,y_ij,z_ij) vectors }
\textcolor{red}{\bf q6mat}: \textcolor{green}{CONTACT_MATRIX} ...
  GROUP=@mdatoms COMPONENTS 
  SWITCH=\{RATIONAL D_0=0.6 R_0=1 NN=6 MM=12 D_MAX=2.0\}
...
\textcolor{blue}{ # Evaluate the Y_l^m values for each bond}
\textcolor{red}{\bf q6sh}: \textcolor{green}{SPHERICAL_HARMONIC} ARG=\textcolor{red}{q6mat.x},\textcolor{red}{q6mat.y},\textcolor{red}{q6mat.z},\textcolor{red}{q6mat.w} L=6
\textcolor{blue}{ # Calculate the vector of q_lm(i) values}
\textcolor{blue}{\bf aones}: \textcolor{green}{ONES} SIZE=@natoms
\textcolor{blue}{\bf q6sp}: \textcolor{green}{MATRIX_VECTOR_PRODUCT} ARG=\textcolor{red}{q6sh.*},\textcolor{blue}{aones}
\textcolor{blue}{ # Evaluate the |q_l(i)| values} 
\textcolor{blue}{\bf q6norm2:} \textcolor{green}{COMBINE} ...
   PERIODIC=NO ARG=\textcolor{blue}{q6sp.*} 
   POWERS=2,2,2,2,2,2,2,2,2,2,2,2,2,2,2,2,2,2,2,2,2,2,2,2,2,2 
...
\textcolor{blue}{\bf q6norm}: \textcolor{green}{CUSTOM} ARG=\textcolor{blue}{q6norm2} FUNC=sqrt(x) PERIODIC=NO
\textcolor{blue}{ # Construct a matrix in which the ith row contains the set of q_lm(i) values} 
\textcolor{red}{\bf vecs}: \textcolor{green}{VSTACK} ARG=\textcolor{blue}{q6sp.*}
\textcolor{blue}{ # Divide each of the q_lm(i) values in the matrix we just calculated by |q_l(i)|} 
\textcolor{blue}{\bf ones:} \textcolor{green}{ONES} SIZE=26
\textcolor{red}{\bf normmat:} \textcolor{green}{OUTER_PRODUCT} ARG=\textcolor{blue}{q6norm},\textcolor{blue}{ones} 
\textcolor{red}{\bf uvecs}: \textcolor{green}{CUSTOM} ARG=\textcolor{red}{vecs},\textcolor{red}{normmat} FUNC=x/y PERIODIC=NO
\textcolor{blue}{ # Calculate a matrix containing all the Q(i,j) values}
\textcolor{blue}{ # Notice that we use a MASK here so Q(i,j) is not calculated}
\textcolor{blue}{ # if atoms i and j are further apart than d_max}
\textcolor{red}{\bf uvecsT}: \textcolor{green}{TRANSPOSE} ARG=\textcolor{red}{uvecs}
\textcolor{red}{\bf dpmat}: \textcolor{green}{MATRIX_PRODUCT} ARG=\textcolor{red}{uvecs},\textcolor{red}{uvecsT} MASK=\textcolor{red}{q6mat.w}
\textcolor{red}{\bf prod}: \textcolor{green}{CUSTOM} ARG=\textcolor{red}{q6mat.w},\textcolor{red}{dpmat} FUNC=x*y PERIODIC=NO
\textcolor{blue}{ # Evaluate the numerator in the expression for s_i above}
\textcolor{blue}{\bf numer}: \textcolor{green}{MATRIX_VECTOR_PRODUCT} ARG=\textcolor{red}{prod},\textcolor{blue}{aones}
\textcolor{blue}{ # Evaluate the coordination numbers}
\textcolor{blue}{\bf denom}: \textcolor{green}{MATRIX_VECTOR_PRODUCT} ARG=\textcolor{red}{q6mat.w},\textcolor{blue}{aones}
\textcolor{blue}{ # These are the s_i values}
\textcolor{blue}{\bf lq6}: \textcolor{green}{CUSTOM} ARG=\textcolor{blue}{numer},\textcolor{blue}{denom} FUNC=x/y PERIODIC=NO
{\bf mean}: \textcolor{green}{MEAN} ARG=\textcolor{blue}{lq6} PERIODIC=NO
 \textcolor{green}{BIASVALUE} ARG=mean
\end{Verbatim}
\end{tcolorbox}

\begin{figure}
    \includegraphics[width=0.6\textwidth]{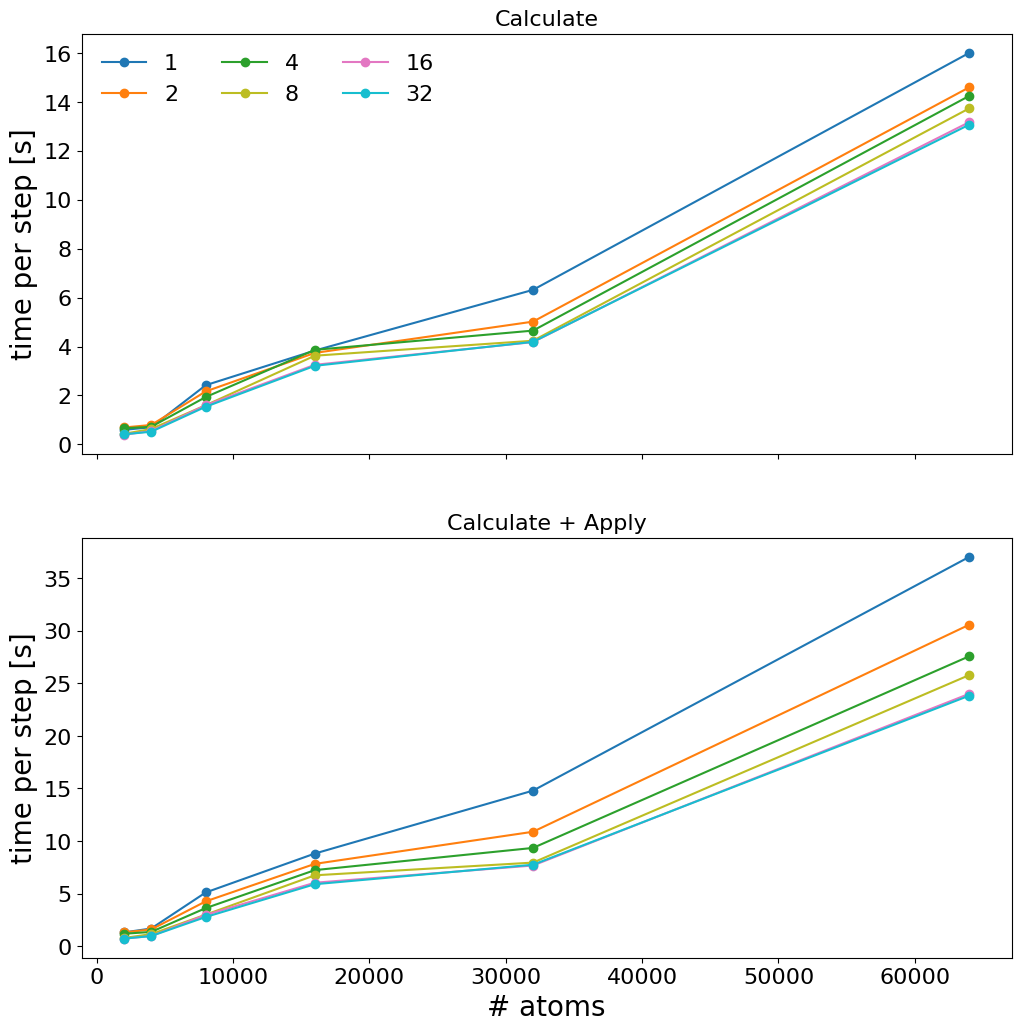}
    \caption{The cost of a single PLUMED step changes as a function of the number of atoms for which the quantity defined in equation \ref{eqn:lq6-proper} is being computed. The top panel shows how the cost of calculating this quantity changes, while the bottom panel shows how the cost of calculating this quantity and applying a force upon it changes. The various lines show the costs when the calculation is run on the numbers of OpenMP threads indicated in the legend.}
    \label{fig:lq6-performance}
\end{figure}

In previous versions of PLUMED the computational expense associated with doing the calculations in the input above was much greater than what is required to compute equation \ref{eqn:q6}.  However, a comparison between Fig. \ref{fig:q6-performance} and \ref{fig:lq6-performance} shows that there is no longer a large difference in the cost associated with computing equations \ref{eqn:q6} and \ref{eqn:lq6-proper}.  In other words, there is no longer a large computational penalty if you compute equation \ref{eqn:lq6-proper} instead of \ref{eqn:q6}.  

Computing equation \ref{eqn:lq6-proper} was expensive in earlier versions of PLUMED because the distances that are used in the {\tt CONTACT\_MATRIX} action were computed twice. These calculations are still done twice if you use the {\tt LOCAL\_Q6} shortcut that is provided within PLUMED to maintain the older syntax for this command. By breaking up the various actions within PLUMED and allowing one to reuse expensive-to-compute values computed in one action in other parts of the input we have also improved the performance of the code.   

\begin{figure}
\centering
\includegraphics[width=0.5\textwidth]{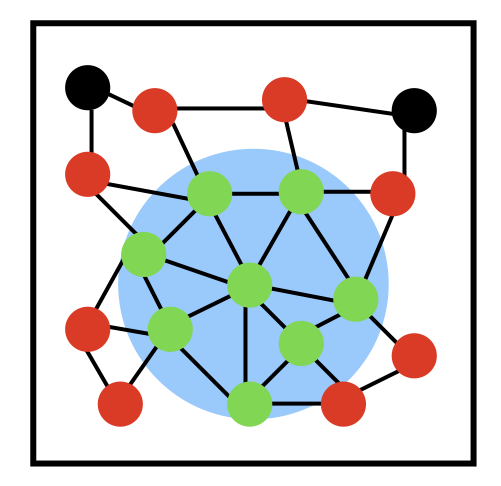}
\caption{Optimizing the evaluation local Q6 parameters (equation \ref{eqn:lq6-proper}) in a sphere. %The process that is used evaluating equation \ref{eqn:lq6-proper} for only those atoms that are within the blue circle.  
The lines indicate pairs of atoms that are within $d_{max}$ of each other.  The green circles are the atoms that are within the blue region and for which we need to evaluate equation \ref{eqn:lq6-proper}. To evaluate equation \ref{eqn:lq6-proper} for these atoms we need to evaluate $q_{lm}$ values for all the atoms that are shown in green and all the atoms shown in red that are within $d_{max}$ of a green atom. Many of the atoms shown in red, for which we need to evaluate $q_{lm}$, are not within the blue circle.}
\label{fig:vol-lq6}
\end{figure}

Now suppose that you want to compute the average value of the quantity that is defined in equation \ref{eqn:lq6-proper} for the subset of atoms that are in a particular part of the box. We cannot use the volume action in the input to the MASK keyword for the {\tt CONTACT\_MAP} action with label \textcolor{red}{\bf cmap} from the above input in the way  that was illustrated in section \ref{sec:volumes} because, as illustrated in Fig. \ref{fig:vol-lq6}, we need to evaluate $q_{lm}(i)$ values for atoms that are not within the region of interest in order to evaluate equation \ref{eqn:lq6-proper} for all the atoms in the region of interest.

We can still use a MASK to reduce the expense of this calculation as is illustrated by the input that is shown below:

\begin{tcolorbox}[top=-10pt]
\singlespacing \footnotesize
\begin{Verbatim}[commandchars=\\\{\}]
\textcolor{blue}{\bf aones}: \textcolor{green}{ONES} SIZE=@natoms
\textcolor{blue}{# Create an atom that is fixed at the origin}
\textcolor{violet}{\bf center}: \textcolor{green}{FIXEDATOM} AT=0,0,0
\textcolor{blue}{# Determine if each of the atoms is within a sphere of radius 1.5 nm that is}
\textcolor{blue}{# centered on the point (0,0,0)}
\textcolor{blue}{\bf w}: \textcolor{green}{INSPHERE} ...
    ATOMS=@mdatoms CENTER=\textcolor{violet}{center} 
    RADIUS=\{RATIONAL D_0=15.9 R_0=0.01 D_MAX=16\}
...
\textcolor{blue}{ # Compute the contact matrix for the atoms that are within the sphere }
\textcolor{red}{\bf lq6mat}: \textcolor{green}{CONTACT_MATRIX} ...
  GROUP=@mdatoms  
  SWITCH=\{RATIONAL D_0=0.6 R_0=1 NN=6 MM=12 D_MAX=2.0\}
  MASK=\textcolor{blue}{w}
...
\textcolor{blue}{ # Construct a matrix in which every row is equal to the vector w}
\textcolor{red}{\bf volmat}: \textcolor{green}{OUTER_PRODUCT} ARG=\textcolor{blue}{w},\textcolor{blue}{aones} MASK=\textcolor{red}{lq6mat}
\textcolor{blue}{ # Element i,j of this matrix is non-zero if r_ij is less than d_max }
\textcolor{blue}{ # and atom i is in the region of interest.}
\textcolor{red}{\bf bondsmat}: \textcolor{green}{CUSTOM} ARG=\textcolor{red}{lq6mat},\textcolor{red}{volmat} FUNC=x*y PERIODIC=NO
\textcolor{blue}{ # Transposing the matrix above and multiplying it by a vector of ones}
\textcolor{blue}{ # results in a vector in which element i is only non zero if there is a bond}
\textcolor{blue}{ # between it and one of the atoms in the region of interest.  }
\textcolor{red}{\bf bondsmatT}: \textcolor{green}{TRANSPOSE} ARG=\textcolor{red}{bondsmat}
\textcolor{blue}{\bf bonds}: \textcolor{green}{MATRIX_VECTOR_PRODUCT} ARG=\textcolor{red}{bondsmatT},\textcolor{blue}{aones}
\textcolor{blue}{ # The set of atoms for which we need to compute q_lm(i) values includes }
\textcolor{blue}{ # the atoms in the region of interest and the the atoms that are bonded}
\textcolor{blue}{ # to atoms in the region of interest.}
\textcolor{blue}{\bf q6mask}: \textcolor{green}{CUSTOM} ARG=\textcolor{blue}{bonds},\textcolor{blue}{w} FUNC=x+y PERIODIC=NO
\textcolor{blue}{ # In the next few lines we use the q6mask value in the input to the MASK}
\textcolor{blue}{ # keyword so we only calculate the q_lm(i)/|q_l(i) values using the method }
\textcolor{blue}{ # described in the previous input for the subset of atoms that are required.}
\textcolor{red}{\bf q6mat}:  \textcolor{green}{CONTACT_MATRIX} ...
  GROUP=@mdatoms  COMPONENTS
  SWITCH=\{RATIONAL D_0=0.6 R_0=1 NN=6 MM=12 D_MAX=2.0\}
  MASK=\textcolor{blue}{q6mask}
...
\textcolor{red}{\bf q6sh}: \textcolor{green}{SPHERICAL_HARMONIC} ARG=\textcolor{red}{q6mat.x},\textcolor{red}{q6mat.y},\textcolor{red}{q6mat.z},\textcolor{red}{q6mat.w} L=6 
\textcolor{blue}{\bf q6sp}: \textcolor{green}{MATRIX_VECTOR_PRODUCT} ARG=\textcolor{red}{q6sh.*},\textcolor{blue}{aones} 
\textcolor{blue}{\bf q6norm2}: \textcolor{green}{COMBINE} ...
   PERIODIC=NO ARG=\textcolor{blue}{q6sp.*} MASK=\textcolor{blue}{q6mask}
   POWERS=2,2,2,2,2,2,2,2,2,2,2,2,2,2,2,2,2,2,2,2,2,2,2,2,2,2 
...
\textcolor{blue}{\bf q6norm}: \textcolor{green}{CUSTOM} ARG=\textcolor{blue}{q6norm2} FUNC=sqrt(x) MASK=\textcolor{blue}{q6mask} PERIODIC=NO
\textcolor{red}{\bf vecs}: \textcolor{green}{VSTACK} ARG=\textcolor{blue}{q6sp.*} 
\textcolor{blue}{\bf ones}: \textcolor{green}{ONES} SIZE=26
\textcolor{red}{\bf normmat}: \textcolor{green}{OUTER_PRODUCT} ARG=\textcolor{blue}{q6norm},\textcolor{blue}{ones} 
\textcolor{red}{\bf uvecs}: \textcolor{green}{CUSTOM} ARG=\textcolor{red}{vecs},\textcolor{red}{normmat} FUNC=x/y PERIODIC=NO
\end{Verbatim}
\end{tcolorbox}
\begin{tcolorbox}[top=-10pt]
\singlespacing \footnotesize
\begin{Verbatim}[commandchars=\\\{\}]
\textcolor{blue}{ # Now that we have the q_lm(i)/|q_l(i)| values we can calculate }
\textcolor{blue}{ # the order parameter of interest using the method that was explained }
\textcolor{blue}{ # in the previous input}
\textcolor{red}{\bf uvecsT}: \textcolor{green}{TRANSPOSE} ARG=\textcolor{red}{uvecs}
\textcolor{red}{\bf dpmat}: \textcolor{green}{MATRIX_PRODUCT} ARG=\textcolor{red}{uvecs},\textcolor{red}{uvecsT} MASK=\textcolor{red}{lq6mat}
\textcolor{red}{\bf prod}: \textcolor{green}{CUSTOM} ARG=\textcolor{red}{lq6mat},\textcolor{red}{dpmat} FUNC=x*y PERIODIC=NO
\textcolor{blue}{\bf numer}: \textcolor{green}{MATRIX_VECTOR_PRODUCT} ARG=\textcolor{red}{prod},\textcolor{blue}{aones}
\textcolor{blue}{\bf denom}: \textcolor{green}{MATRIX_VECTOR_PRODUCT} ARG=\textcolor{red}{lq6mat},\textcolor{blue}{aones}
\textcolor{blue}{ # These are the s_i values so we use w in the input for MASK}
\textcolor{blue}{\bf lq6}: \textcolor{green}{CUSTOM} ARG=\textcolor{blue}{w},\textcolor{blue}{numer},\textcolor{blue}{denom} FUNC=x*y/z PERIODIC=NO
\textcolor{blue}{ # Sum of the order parameter for atoms in the region of interest}
{\bf opsum}: \textcolor{green}{SUM} ARG=\textcolor{blue}{lq6} PERIODIC=NO
\textcolor{blue}{ # Number of atoms in region of interest}
{\bf natoms}: \textcolor{green}{SUM} ARG=\textcolor{blue}{w} PERIODIC=NO
\textcolor{blue}{ # Average value of order parameter in region of interest}
{\bf mean}: \textcolor{green}{CUSTOM} ARG=opsum,natoms FUNC=x/y PERIODIC=NO
 \textcolor{green}{BIASVALUE} ARG=mean
\end{Verbatim}
\end{tcolorbox}

\begin{figure}
\includegraphics[width=0.8\textwidth]{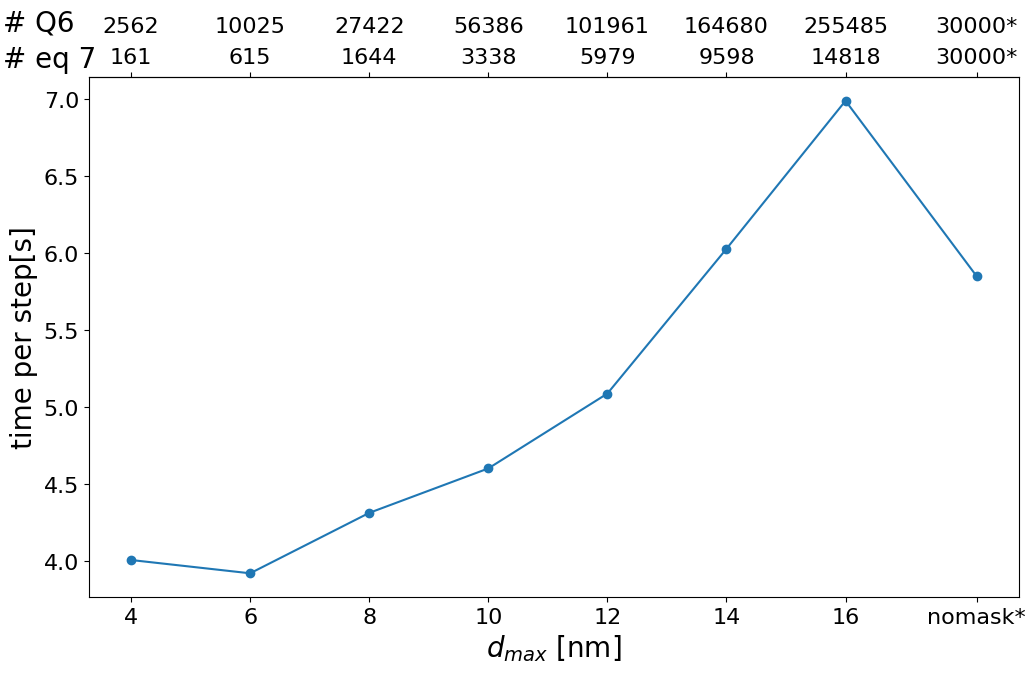}
\caption{The cost of a single PLUMED step as a function of the radius of the spherical region in which the quantity defined in equation \ref{eqn:lq6} is being calculated. The bottom $x$-axis indicates the radius of the spherical region in which equation \ref{eqn:lq6} is being evaluated, while the numbers labeled \#eq 7 on the top $x$ axis indicate the number of atoms that are within it. The numbers labeled \# Q6 are the number of atoms for which equation \ref{eqn:q6} must be evaluated.}
\label{fig:lq6-vol-performance}
\end{figure}

As you have to evaluate two {\tt CONTACT\_MATRIX} actions in the above input there will be cases where simply computing the values of $s_i$ using equation \ref{eqn:lq6-proper} for all the atoms in the system and then computing the average value of this quantity in the region of interest without using the MASK keyword at all is computationally cheaper than using the input above.  We thus ran calculations to determine how
%when the above input is cheaper, Fig. \ref{fig:lq6-vol-performance} shows how 
the length of time required to perform the calculation using 16 OpenMP threads for a system of 30000 atoms depends on the radius of the spherical region (Fig. \ref{fig:lq6-vol-performance}).  The bottom $x$-axis in this figure indicates the radius of the sphere in which equation~\ref{eqn:lq6} is being evaluated. The top axis then shows how many atoms we need to evaluate equation~\ref{eqn:q6} for in order to evaluate equation~\ref{eqn:lq6} for the atoms in the spherical region as well as the number of atoms in the sphere.  The result is as you would expect; namely, when the sphere is large the computational expense associated with calculating the two contact matrices ensures that using the input above is slower than simply calculating equation~\ref{eqn:q6} for all atoms and averaging. However, when the sphere is smaller the reduction in computational cost that is associated with evaluating equation~\ref{eqn:q6} for a smaller number of atoms easily makes up for the cost that is added by evaluating the contact matrix twice. It is thus considerably more computationally efficient to use the input above whenever the volume of interest is small. 

\subsection{Additional tips}

Having discussed some of the most recent optimizations added to the PLUMED codebase, we here report  a checklist of other performance-optimization ideas that have been implemented and, that are discussed in previous papers or online tutorials:
\begin{itemize}
    \item When using metadynamics you should employ the implementation of the bias that stores the potential on a grid.  You do this by employing the {\tt GRID\_MIN}, {\tt GRID\_MAX} and {\tt GRID\_BIN} keywords in your {\tt METAD} command as discussed in \cite{ptutorial07}.
    \item At the time of writing, some collective variables in PLUMED use a standard neighbor list rather than the linked list strategy. A notable example is {\tt COORDINATION}. Typically the keywords {\tt NL\_CUTOFF} and {\tt NL\_STRIDE} are used to turn on the neighbor list. If you want to use this feature you will need to optimize the neighbor list parameters for both speed and correctness as discussed in \cite{ptutorial07}.
    \item Some biasing potentials act on collective variables that have a smooth dependence on the atomic coordinates.  When using such bias potentials you can use the multiple-time-step implementation discussed in \cite{bussi-nstep,ptutorial07}.
    \item Some actions in PLUMED are able to modify global coordinates. Examples include {\tt WHOLEMOLECULES} and {\tt FIT\_TO\_TEMPLATE}. For these actions, PLUMED cannot track dependencies in an optimal way. This means that you should carefully choose the {\tt STRIDE} parameters for these actions to have correct results and to minimize their impact on the overall performances.
    \item The benchmark tool described above is very useful for comparing PLUMED versions and different input files. However, real world performances might depend on technicalities related to the underlying MD code calculations, such as transfer of atoms from/to the GPU, use of caches, etc. It is always recommended to fine tune your input files in an as-realistic-as-possible scenario, which often means running a short version of your production trajectory.
\end{itemize}

\section{Conclusion}

%% MAX draft of conclusions following Giovanni's comments
In this tutorial, we have demonstrated how to perform reliable and reproducible benchmarks of PLUMED’s performance using the recently introduced {\tt plumed benchmark} tool. We have used this tool to present a series of benchmarks covering a diverse set of applications, from simple scalar quantities to more complex collective variables such as symmetry functions and Steinhardt parameters. These examples illustrate how performance can be optimized by employing vector-based operations, linked-list algorithms, and appropriate parallelization strategies.

We encourage developers who contribute new functionalities to PLUMED to follow a similar benchmarking approach. Providing benchmarks alongside contributed code not only helps ensure performance portability and transparency but also facilitates meaningful comparisons between implementations across different hardware and software environments. We also invite developers to explore alternative implementations of the vectorized calculations discussed here, for instance using emerging numerical frameworks such as JAX \cite{pytorch}, PyTorch \cite{pytorch}, or TensorFlow \cite{tensorflow}, and to report and share their benchmarking results with the community. Additional development work and careful benchmarking would likely result in further improvements for all the methods discussed in this tutorial.  Our hope is that by providing sufficient detail for readers to re-implement these functionalities elsewhere and benchmark new implementations against our own, we embrace the competitive and collaborative spirit that has always driven the best scientific software development — one that values both innovation and rigorous evaluation in equal measure.

Looking forward, we envision that benchmarking could be further integrated into PLUMED’s development workflow. Automated benchmarking pipelines could regularly assess performance across multiple PLUMED versions and hardware configurations, generating plots similar to those shown here and enabling continuous monitoring of performance evolution. Such a system would not only streamline performance testing but also strengthen PLUMED’s role as a transparent and reproducible platform for method development in molecular simulations.

\begin{acknowledgement}

M.B. would like to acknowledge funding from the European Research Council (ERC) under
the European Union’s Horizon 2020 research and innovation programme (Grant agreement
No. 101086685 – bAIes).
D.R. and G.B. acknowledge the Italian National Centre for HPC, Big Data, and Quantum Computing (grant No. CN00000013), founded within the Next Generation EU initiative.
B. Tribello is thanked for drawing the image for the table of contents. Lastly, all the authors thank Chris Chipot for the  patience he demonstrated in editing this paper.   

\end{acknowledgement}

%%%%%%%%%%%%%%%%%%%%%%%%%%%%%%%%%%%%%%%%%%%%%%%%%%%%%%%%%%%%%%%%%%%%%
%% The same is true for Supporting Information, which should use the
%% suppinfo environment.
%%%%%%%%%%%%%%%%%%%%%%%%%%%%%%%%%%%%%%%%%%%%%%%%%%%%%%%%%%%%%%%%%%%%%
\begin{suppinfo}

Some supporting information

\end{suppinfo}

%%%%%%%%%%%%%%%%%%%%%%%%%%%%%%%%%%%%%%%%%%%%%%%%%%%%%%%%%%%%%%%%%%%%%
%% The appropriate \bibliography command should be placed here.
%% Notice that the class file automatically sets \bibliographystyle
%% and also names the section correctly.
%%%%%%%%%%%%%%%%%%%%%%%%%%%%%%%%%%%%%%%%%%%%%%%%%%%%%%%%%%%%%%%%%%%%%
\bibliography{achemso-demo}

\end{document}